\documentclass[twocolumn,amsmath,amssymb,prd]{revtex4}

\usepackage{epsfig}

\def\al{\alpha}
\def\be{\beta}

\def\de{\delta}

\def\ve{\varepsilon}

\def\th{\theta}

\def\la{\lambda}

\def\si{\sigma}

\def\ta{\tau}

\def\ch{\chi}

\def\De{\Delta}

\def\fr#1#2{{{#1} \over {#2}}}
\def\frac#1#2{{\textstyle{{#1} \over {#2}}}}
\def\half{{\textstyle{1\over 2}}}
\def\ol{\overline}

\def\lsim{\mathrel{\rlap{\lower4pt\hbox{\hskip1pt$\sim$}}
    \raise1pt\hbox{$<$}}}
\def\gsim{\mathrel{\rlap{\lower4pt\hbox{\hskip1pt$\sim$}}
    \raise1pt\hbox{$>$}}}

\def\etal{{\it et al.}}

\newcommand{\beq}{\begin{equation}}
\newcommand{\eeq}{\end{equation}}
\newcommand{\bea}{\begin{eqnarray}}
\newcommand{\eea}{\end{eqnarray}}
\newcommand{\rf}[1]{(\ref{#1})}
\newcommand{\bM}{\begin{pmatrix}}
\newcommand{\eM}{\end{pmatrix}}

\def\nn{\nonumber}

\def\ring#1{{\mathaccent'27 #1}}
\def\ari{{\ring{a}}}
\def\cri{{\ring{c}}}
\def\kri{{\ring{k}}}

\def\a#1{\ari^{(#1)}}
\def\C#1{\cri^{(#1)}}
\def\k#1{\kri^{(#1)}}

\def\msm{3$\nu$SM} 

\def\h{h^\nu_{\rm eff}}
\def\hp{h'^{\nu}_{\rm eff}}
\def\hb{h^{\overline\nu}_{\rm eff}}

\def\lto{L_{31}}
\def\ltob{\ol L_{31}}
\def\ltt{L_{21}}
\def\lttb{\ol L_{21}}

\def\nub{\ol\nu}
\def\mix{\leftrightarrow}

\def\ceafm{$c_8a_5m$}
\def\ceafcfm{$c_8a_5c_4m$}
\def\cfatm{$c_4a_3m$}
\def\cscfm{$c_6c_4m$}

\begin{document}
\title{Lorentz- and CPT-violating models for neutrino oscillations}

\author{Jorge S.\ D\'\i az and V.\ Alan Kosteleck\'y}

\affiliation{Physics Department, Indiana University, 
Bloomington, IN 47405, U.S.A.}

\date{IUHET 561, August 2011}

\begin{abstract}

A class of calculable global models for neutrino oscillations 
based on Lorentz and CPT violation is presented.
One simple example matches established neutrino data
from accelerator, atmospheric, reactor, and solar experiments, 
using only two degrees of freedom instead of the usual five.
A third degree of freedom appears in the model,
and it naturally generates the MiniBooNE low-energy anomalies.
More involved models in this class can also accommodate 
the LSND anomaly and neutrino-antineutrino differences
of the MINOS type.
The models predict some striking signals
in various ongoing and future experiments.

\end{abstract}

\maketitle

\section{Introduction}

The minimal Standard Model (SM) of particle physics
contains three flavors of massless left-handed neutrinos.
However,
experiments with solar, reactor, accelerator, 
and atmospheric neutrinos have convincingly demonstrated 
the existence of neutrino flavor oscillations.
This effect cannot be accommodated within the SM
and so represents forceful evidence for new physics.

A popular hypothesis attributes neutrino oscillations
to the existence of a tiny neutrino mass matrix
with off-diagonal components.
Extending the SM to incorporate this notion 
produces a model with three flavors 
of massive neutrinos (\msm),
in which oscillations are controlled by a 3$\times$3 matrix
involving six parameters:
two mass-squared differences
$\De m^2_\odot$, $\De m^2_\text{atm}$,
three angles
$\th_{12}$, $\th_{23}$, $\th_{13}$, 
and a phase $\de$ controlling CP violation. 
The first four of these parameters must be nonzero
to match established experimental data,
while recent results provide indications that 
the angle $\th_{13}$ must also be nonzero
\cite{pdg,t2krecent}.

In this work,
we explore an alternative hypothesis 
attributing part of the observed neutrino oscillations
to tiny Lorentz and CPT violation,
which might arise in a Planck-scale theory 
unifying gravity and quantum physics
such as string theory
\cite{ksp}.
One motivation for studying alternative hypotheses 
for neutrino oscillations is based on existing data.
Several neutrino experiments have reported 
potential evidence for anomalous neutrino oscillations
that is incompatible with the \msm.
This includes the LSND signal 
\cite{LSND}, 
the MiniBooNE low-energy excess 
\cite{MiniBooNE1}, 
and neutrino-antineutrino differences 
in the MiniBooNE 
\cite{MiniBooNE2} 
and MINOS 
\cite{MINOSanomaly}
experiments.
Another motivation is philosophical:
having more than one viable hypothesis
is known to be of great value 
in guiding experimental and theoretical investigations
of new physics.
Lorentz and CPT violation is interesting in this context 
because it naturally generates neutrino oscillations
and moreover leads to simple global models
describing all established and anomalous neutrino data
\cite{km,puma}.

An appropriate theoretical framework for studying 
realistic signals of Lorentz violation is 
effective field theory 
\cite{kp}.
In this context,
CPT violation is necessarily accompanied
by Lorentz violation
\cite{owg},
and the comprehensive description 
for Lorentz and CPT violation 
containing the SM and General Relativity
is given by the Standard-Model Extension (SME)
\cite{ck,akgrav}.
In the SME action,
each Lorentz-violating term 
is a coordinate-independent quantity
constructed from the product of a Lorentz-violating operator
and a controlling coefficient.
The combination of observer coordinate invariance 
and Lorentz violation
implies particles in the SME follow trajectories
in a pseudo-Riemann-Finsler geometry 
\cite{finsler}.

Over the last decade or so,
many experimental analyses using a broad variety of techniques
have been performed
to seek nonzero SME coefficients for Lorentz and CPT violation
\cite{tables}.
The interferometric nature of particle oscillations
suggests that sensitive neutrino or neutral-meson experiments 
might well yield the first detectable signals
of tiny Lorentz violation.
In the neutrino sector,
recent SME-based phenomenological studies
\cite{km,puma,bicycle,scsk,gl,tandem,%
nusme1,nusme2,nusme3,nusme4,nusme5,nusme6,nusme7,%
nusme8,nusme9,nusme10,nusme11,nusme12,nusme13,nusme14}
and methodologies for experimental analysis 
\cite{kmshortbaseline,kmd} 
have spurred searches 
for Lorentz and CPT violation by
the LSND
\cite{lsndlv}, 
Super-Kamiokande (SK)
\cite{sklv},
MINOS
\cite{minoslv1,minoslv2},
MiniBooNE
\cite{MiniBooNElv},
and IceCube collaborations 
\cite{IceCube}.
Searches have also been performed with neutral mesons
\cite{ak,mesons},
and recent D0 results suggest 
some evidence for anomalous CP violation
\cite{d0}
that could be attributed to Lorentz and CPT violation
\cite{kvk}.

Here,
we focus on a special class of `puma' models
in which the $3\times 3$ effective hamiltonian $\h$ 
governing oscillations
of three flavors of active left-handed neutrinos 
is characterized by two simple properties:
isotropic Lorentz violation,
and a zero eigenvalue
\cite{puma}.
The isotropic Lorentz violation
implies boost invariance is broken
while leaving rotations unaffected,
so $\h$ is independent of the direction 
of the neutrino momentum 
but must contain unconventional dependence 
on the neutrino energy $E$.
This leads to unconventional energy dependences
even in vacuum oscillations,
producing a broad range of unique neutrino behavior. 
The zero eigenvalue can be attributed 
to a discrete symmetry of $\h$.
It ensures quadratic calculability 
of the mixing matrix and of oscillation probabilities
for all models,
even when matter effects are included.
These two features differ qualitatively
from the \msm,
in which the Lorentz-invariant mass terms
force a $1/E$ energy dependence of all terms in $\h$
and the lack of symmetry results in calculational complexity. 

The unconventional energy dependence in $\h$
generically takes the form of polynomials in $E$
arising from Lorentz-violating operators of arbitrary dimension 
in the SME Lagrange density
\cite{kmnonmin}.
The polynomial coefficients are therefore determined 
in terms of SME coefficients for Lorentz violation.
For much of this work
we make the plausible assumption 
that a few terms of comparatively low mass dimension dominate
the neutrino behavior, 
either by chance
or due to the presently unknown structure
of the underlying theory,
and hence that only a few coefficients are needed
to reproduce the bulk of existing neutrino data. 
Indeed,
the basic puma models considered below
have only three degrees of freedom,
which includes one mass 
and two Lorentz-violating coefficients.
Remarkably,
two of these degrees of freedom
suffice to reproduce all established neutrino behavior,
a frugal result compared to 
the five degrees of freedom required by the \msm.
Moreover,
the third degree of freedom
naturally reproduces the anomalous results found by MiniBooNE 
\cite{MiniBooNE1,MiniBooNE2} 
without introducing new particles or forces. 
Comparatively minor modifications 
of these simple puma models
that preserve the discrete symmetry of $\h$ 
can also accommodate 
the LSND signal 
\cite{LSND}
and anomalies of the MINOS type 
\cite{MINOSanomaly}.

The structure of this paper is as follows.
The basic properties of the general puma models
are presented in Sec.\ \ref{Sec: puma_model}.
Applications to existing experiments
are discussed in Sec.\ \ref{Experiments}.
A specific model involving one mass parameter 
and two Lorentz-violating operators,
one of which is CPT odd, 
is used for illustrative purposes.
Predictions for future experiments
are presented in Sec.\ \ref{Predictions}.
Some of these are strikingly different 
from models based on the \msm.
Variant puma models using three different degrees of freedom
or more than three parameters
are considered in Sec.\ \ref{Variant puma models}.
Finally,
Sec.\ \ref{Discussion}
contains some comments on the general nature of the models.

The notation adopted here is that of Refs.\ \cite{km,puma}.
A mass parameter is denoted $m$,
a coefficient for isotropic CPT-odd Lorentz violation 
is denoted $\a{d}$,
and a coefficient for isotropic CPT-even Lorentz violation 
is denoted $\C{d}$,
where $d$ is the dimension of the corresponding operator.
To identify the various specific puma models
according to their coefficient content,
we introduce a convenient nomenclature
listing coefficients 
in descending order of operator mass dimension.
For example,
a model with three degrees of freedom 
including a mass term $m$
and coefficients $\a{5}$ and $\C{8}$
for Lorentz violation 
is called a \ceafm\ model.

\section{General model}
\label{Sec: puma_model}

In the general puma model,
the effective $3\times3$ hamiltonian $\h$
describing the oscillation of three active neutrino flavors 
$e$, $\mu$, $\ta$ takes the form
\cite{puma}
\beq
\h=A\bM
1 & 1 & 1 \\
1 & 1 & 1 \\
1 & 1 & 1 
\eM
+
B\bM
1 & 1 & 1 \\
1 & 0 & 0 \\
1 & 0 & 0 
\eM
+
C\bM
1 & 0 & 0 \\
0 & 0 & 0 \\
0 & 0 & 0 
\eM,
\label{h(puma)}
\eeq
where $A(E)$, $B(E)$, and $C(E)$ are real functions 
of the neutrino energy $E$.
In this work,
the function $A$ is chosen to be $A=m^2/2E$,
where $m$ is the unique neutrino mass parameter in the theory. 
The functions $B$ and $C$ have nonstandard energy dependence,
which here is taken to arise from Lorentz-violating terms in the SME,
some of which may lie in the nonrenormalizable sector.
The treatment of possible contributions 
to $\h$ from Lorentz-invariant operators 
lies outside our present scope and will be given elsewhere.
We assume all SME coefficients contributing to $\h$
are spacetime constants,
so the model \rf{h(puma)}
incorporates translation invariance
and conserves energy and momentum.
In the context of spontaneous Lorentz violation,
where the SME coefficients can be interpreted 
in terms of expectation values in an underlying theory,
this assumption implies 
soliton solutions, massive modes,
and Nambu-Goldstone modes
\cite{ng}
are disregarded.
The latter may play the role 
of the graviton
\cite{cardinal},
the photon in Einstein-Maxwell theory 
\cite{bumblebee},
or various new forces 
\cite{newforces}.
For simplicity in most specific models considered here,
$B$ and $C$ are taken to be monomials in $E$,
although more complicated polynomials
or nonpolynomial functions can also be of interest.

The function $A$ decreases inversely with energy, 
while $B$ and $C$ typically increase.
At low energies, 
the effective hamiltonian $\h$
is therefore well approximated by the $A$ term alone.
This term has a `democratic' form,
exhibiting symmetry under the permutation group $S_3$
acting on the three neutrino flavors $e$, $\mu$, $\ta$. 
In contrast,
the nonstandard energy dependences
in the $B$ and $C$ terms 
dominate at high energies. 
The flavor-space structure of these terms 
breaks the $S_3$ symmetry to its $S_2$ subgroup 
in the $\mu$-$\ta$ sector.

For antineutrinos,
oscillations are governed by the CPT image $\hb$
of the effective hamiltonian $\h$.
The effect of the CPT transformation on $\h$ 
is to change the signs of any coefficients for Lorentz violation
that are associated with CPT-odd operators in the SME. 
Since mass terms are invariant under CPT
\cite{owg},
the $A$ term in $\h$ is unaffected by the transformation.
At low energies,
the full permutation symmetry of the puma model 
is therefore $S_3\times\ol S_3$, 
where $\ol S_3$ is the symmetry 
acting on antineutrino flavors.
At high energies,
the $S_3\times\ol S_3$ invariance breaks to $S_2\times\ol S_2$.
If any coefficients for CPT-odd Lorentz violation are present,
differences between neutrinos and antineutrinos can become manifest.

An elegant feature of the puma model
is the existence of a zero eigenvalue
for the effective hamiltonian,
which is a consequence of the permutation symmetry 
of the texture \rf{h(puma)}.
This implies considerable calculational simplification 
compared to the \msm\ and typical other neutrino-oscillation models.
Many results can be obtained exactly by hand 
even when all three neutrino flavors mix.
A short calculation reveals that the eigenvalues $\la_{a'}$,
$a' = 1,2,3$, 
of the effective hamiltonian $\h$ take the exact form
\bea
\la_1 &=&
\half\left[3A+B+C - \sqrt{(A-B-C)^2+8(A+B)^2}\right],
\nn\\
\la_2 &=&
\half\left[3A+B+C + \sqrt{(A-B-C)^2+8(A+B)^2}\right],
\nn\\
\la_3 &=& 0.
\label{las(puma)}
\eea
The mixing matrix $U_{a'a}$
that diagonalizes $\h$ can also be expressed exactly as 
\beq
U_{a'a}=\bM
\dfrac{\la_1-2A}{N_1} & \dfrac{A+B}{N_1} & \dfrac{A+B}{N_1} \\\\
\dfrac{\la_2-2A}{N_2} & \dfrac{A+B}{N_2} & \dfrac{A+B}{N_2} \\\\
0 & -\dfrac{1}{\sqrt2} & \dfrac{1}{\sqrt2} 
\eM.
\label{U}
\eeq
In this equation,
the index $a$ ranges over $a=e,\mu,\ta$
and the normalization factors are
\bea
N_1&=&\sqrt{(\la_1-2A)^2+2(A+B)^2},
\nn\\
N_2&=&\sqrt{(\la_2-2A)^2+2(A+B)^2}.
\label{normalization}
\eea
The eigenvalues $\ol\la_{a'}$,
the mixing matrix $\ol U_{a'a}$,
and the normalization factors
$\ol N_1$, $\ol N_2$
for the antineutrino effective hamiltonian 
$\hb$ are obtained by CPT conjugation of $B$ and $C$. 

In the low-energy limit,
the mixing matrix \rf{U} reduces to the tribimaximal form
originally postulated on phenomenological grounds
by Harrison, Perkins, and Scott
\cite{hps}.
The democratic structure of the $A$ term in $\h$
therefore ensures tribimaximal mixing 
of the three neutrino flavors at low energies.
Combined with the choice $A = m^2/2E > 0$,
this mixing guarantees agreement of the puma model
with low-energy solar neutrinos 
\cite{Borexino} 
and with the mixing observed in KamLAND 
\cite{KamLAND}. 
For a suitable choice of mass parameter $m$,
as discussed in the next section,
the $A$ term can also correctly describe
the $L/E$ oscillation signature observed by KamLAND 
\cite{KamLAND(L/E)}. 

Another defining feature of the puma model
is a Lorentz-violating seesaw
\cite{km}
that mimics a mass term at high energies,
without invoking mass.
This differs from the usual seesaw mechanism
\cite{seesaw,seesawreview}, 
which is based on mass terms in the action.
Suppose $B$ and $C$ are monomials of the form 
\beq
B(E)=\k{p}E^{p-3},
\quad
C(E)=\C{q}E^{q-3},
\label{B,C(E)}
\eeq
where $p$ and $q$ are the dimensions of the operators 
associated with the coefficients $\k{p}$ and $\C{q}$.
In this work, 
we take $\C{q}>0$ for definiteness
but consider both sign options for $\k{p}$.
Reversing the sign of $\C{q}$ produces phenomenology
closely related to reversing instead the sign of $\k{p}$,
as can be seen by inspecting
Eqs.\ \rf{las(puma)} and \rf{U}.
If $q>p$ then $C$ grows faster than $B$,
so at high energies
\bea
\la_1&\approx&
-\fr{2B^2}{C} 
= -\fr{2(\k{p})^2E^{2p-q-3}}{\C{q}}.
\label{pqlim}
\eea
For the choice $q=2(p-1)$,
the eigenvalue $\la_1$ is proportional to $1/E$
and therefore plays the role of an effective mass term,
even though no mass parameter is present at high energies.
Note that imposing this choice requires 
the dominant coefficient in $C$ to be CPT even. 
The null entries in the $\mu$-$\ta$ block of $\h$ 
and the fast-growing $ee$ element 
guarantee maximal $\nu_\mu\mix\nu_\ta$ mixing
at high energies, 
consistent with observations of atmospheric neutrinos 
\cite{SK,K2K,MINOSnu2011}.
For a suitable choice of the ratio $B^2/C$,
as discussed in the next section,
the seesaw mechanism also reproduces 
the $L/E$ oscillation signature 
in the SK experiment
\cite{SK(L/E)}.

Since the elements of $\h$ are real, 
the probability $P_{\nu_b\to\nu_a}$
of oscillation from $\nu_b$ to $\nu_a$
can be written in the simple form
\beq
P_{\nu_b\to\nu_a}=
\de_{ab}
-4\sum_{a'>b'} 
U_{a'a}U_{a'b}U_{b'a}U_{b'b}
\sin^2(\De_{a'b'}L/2),
\label{probs}
\eeq
where the quantities $\De_{a'b'}= \la_{a'} - \la_{b'}$
are the eigenvalue differences
and $L$ is the baseline. 
For each flavor pair $a$, $b$,
the above sum contains three terms
labeled by the values of $a'$, $b'<a'$.
Each term is the product of an amplitude $-4UUUU$ 
with a sinusoidal phase. 
The antineutrino-oscillation probabilities
$P_{\ol\nu_b\to\ol\nu_a}$
are obtained by CPT conjugation.
Since $A$, $B$, and $C$ are real, 
all processes are T invariant.
As a result,
CP violation occurs if and only if CPT violation does.
Notice that CP-violating effects can appear
even though no analogue of the phase $\de$ in the \msm\
exists in the puma model.

All the above properties 
are insensitive to the $ee$ component of the $B$ term in $\h$. 
As a result,
a modified texture $\hp$ can be constructed 
in which the $ee$ entry in the $B$ term vanishes.
We have verified that most of the properties
discussed in the remainder of this work
remain unchanged for this modified texture. 
One exception is the renormalizable model
presented in Sec.\ \ref{Sec:model_c4a3m},
for which we use a zero $ee$ entry in the $B$ term
because the nonzero value 
produces a tension between the descriptions 
of long-baseline reactor and of solar neutrinos.

\section{Experiments}
\label{Experiments}

Next,
we study the implications of the general model \rf{h(puma)} 
for different experiments.
Many characteristics of the model are generic.
For definiteness,
in this section we illustrate the discussion
with a specific \ceafm\ model 
\cite{puma}.
Some comments on variant models
are provided in Sec.\ \ref{Predictions}.

The numerical values of the three parameters 
in the \ceafm\ model are 
\bea
m^2 &=& 2.6\times10^{-23} {\rm ~GeV}^2, 
\nn\\
\a{5} &=& -2.5\times10^{-19} {\rm ~GeV}^{-1}, 
\nn\\
\C{8} &=& 1.0\times10^{-16} {\rm ~GeV}^{-4},
\label{c8a5m}
\eea
The nonzero value of $\a 5$ implies
this model contains CPT violation.
The value for $m^2$ is consistent
with limits from direct mass measurements
and cosmological bounds
\cite{pdg}.

By construction,
$\a{5}$ and $\C{8}$ are the only nonzero SME coefficients
defined in an isotropic frame $I$.
In some scenarios,
it is reasonable to identify $I$
with a universal inertial frame $U$ 
such as that defined by the cosmic microwave background (CMB),
but other possibilities exist.
Whatever the choice for $I$, 
the experiment frame $E$ is boosted in it
by some combination of the
Earth's motion relative to the CMB,
the Earth's revolution about the Sun,
and the Earth's rotation.
The coefficients $\a{5}$ and $\C{8}$
therefore induce anisotropic effects
via the net boost in $I$.
These could,
for example,
be detected by searches for sidereal or annual variations in $E$
\cite{ak}.
Experimental constraints and signals must be reported
in a specified frame,
but the frame $E$ itself is inappropriate 
because it is noninertial and experiment specific.
By convention,
the canonical inertial frame used to report results 
is a Sun-centered frame $S$ 
\cite{tables,sunframe}.
Inspection reveals that the size of the effects in $S$
induced by the values \rf{c8a5m}
all lie below the sensitivity levels
achieved in experiments to date
\cite{lsndlv,MiniBooNElv,minoslv1,minoslv2,IceCube}.
Future experiments might offer improved sensitivity
and thereby provide a distinct avenue
for testing the model.

\begin{figure}
\begin{center}
\centerline{\psfig{figure=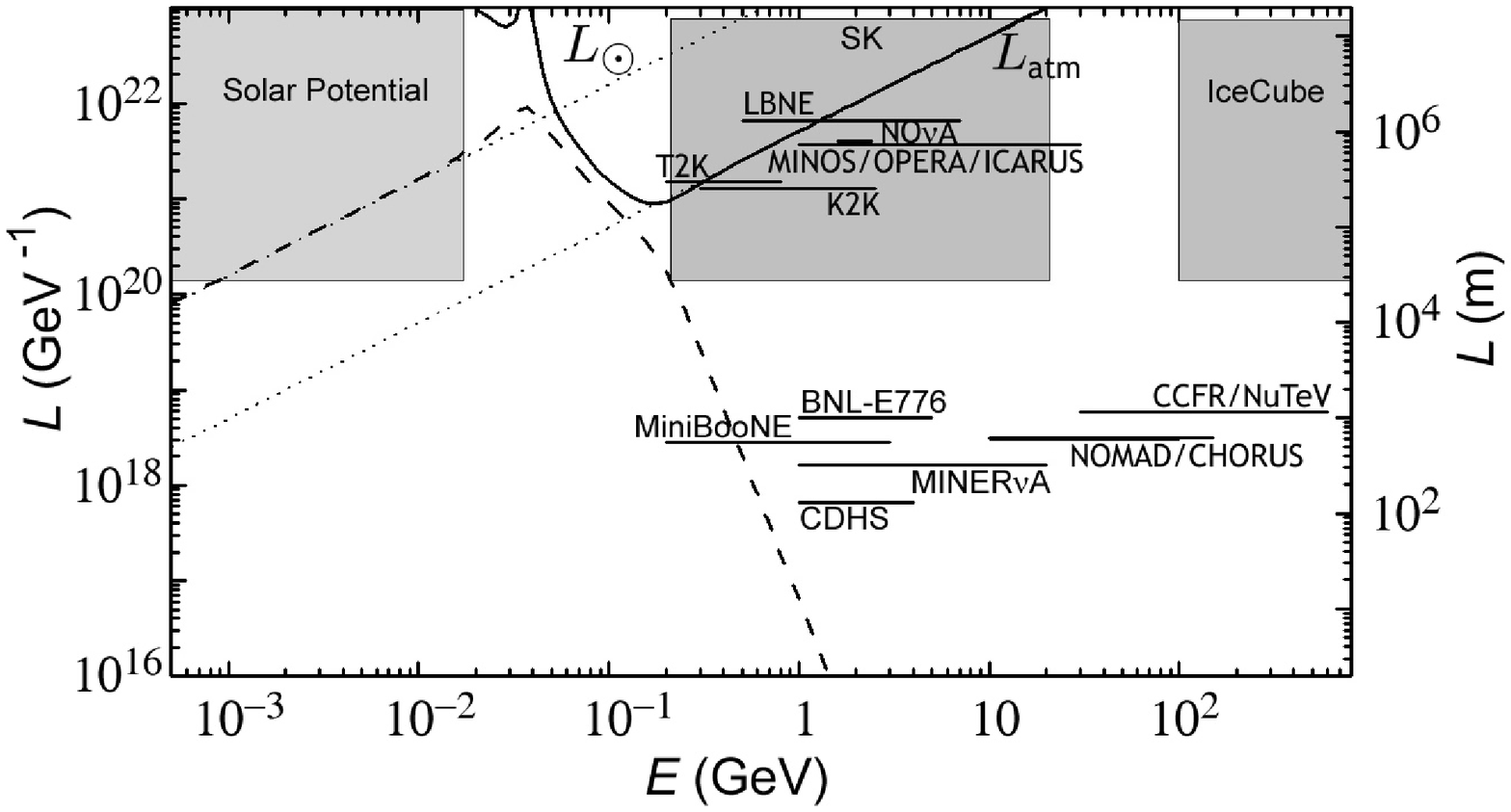,
width=\hsize}}
\centerline{\psfig{figure=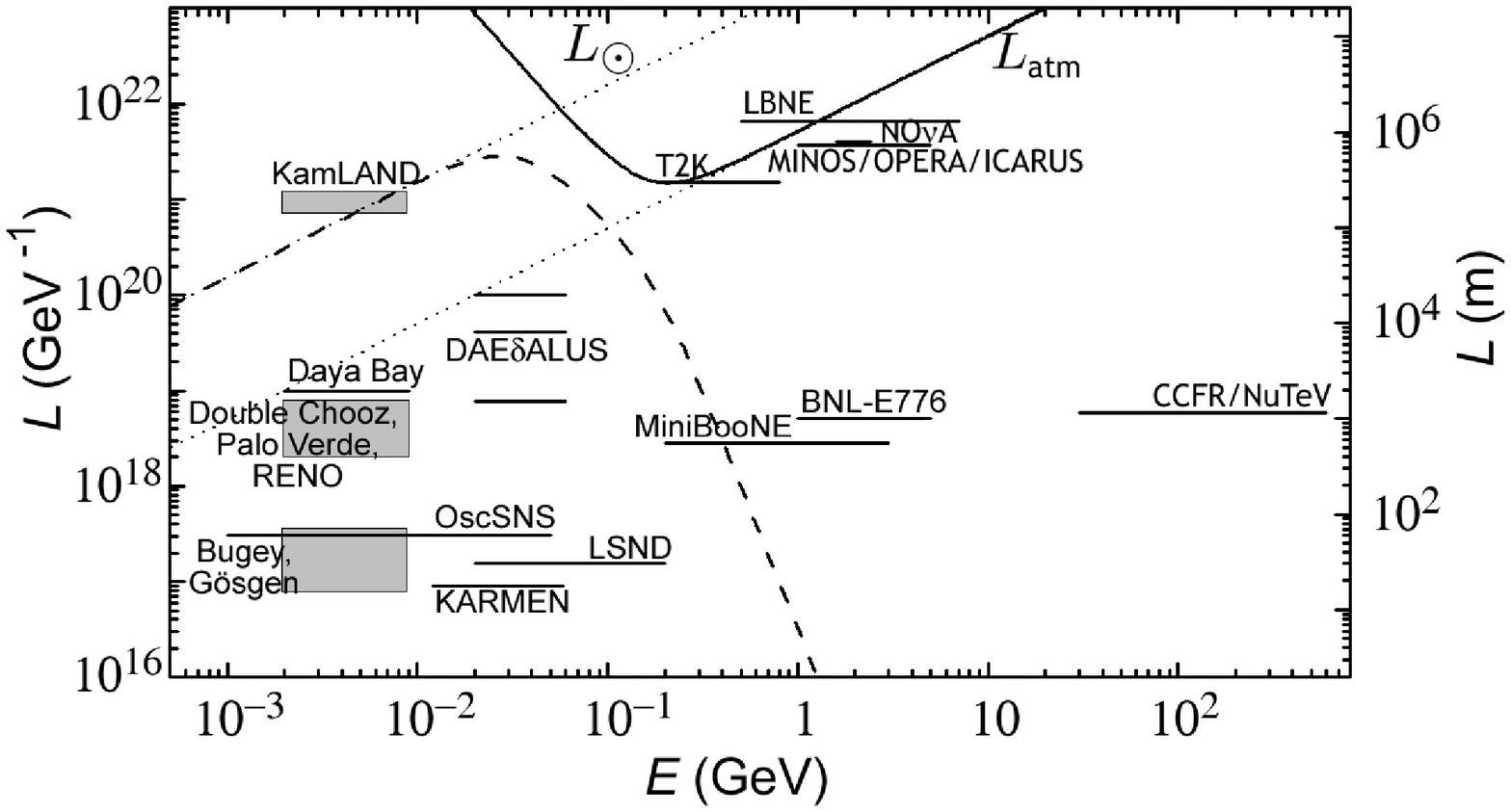,
width=\hsize}}
\caption{ 
Energy dependences of the oscillation lengths
for neutrinos (top) and antineutrinos (bottom).
The disappearance lengths for the puma model are 
$\lto$ (top, solid line), 
$\ltt$ (top, dashed line),
$\ltob$ (bottom, solid line), 
and $\lttb$ (bottom, dashed line),
displayed for the values \rf{c8a5m}.
The dotted lines are the disappearance lengths 
$L_\odot$ (solar) and $L_\text{atm}$ (atmospheric)
in the \msm.
}
\label{kmplot}
\end{center}
\end{figure}

\subsection{General features}

The predictions of any model 
for neutrino and antineutrino oscillations 
can be visualized using a certain plot in $E$-$L$ space 
\cite{km}. 
Experiments are represented on the plot
as regions determined by their baseline and energy coverage,
while a given theory is represented 
by its characteristic oscillation wavelengths
$L_{a'b'} = 2 \pi/|\De_{a'b'}|$
associated with the eigenvalue differences $\De_{a'b'}(E)$.
The absolute value is used 
because the oscillation phase is insensitive 
to the sign of $\De_{a'b'}$.
Each curve $L_{a'b'} = L_{a'b'}(E)$
indicates the first maximum of a kinematic phase 
in the oscillation probability,
thereby establishing the minimal distance 
from the neutrino source required for 
appearance or disappearance signals
in a specific oscillation channel.
Substantial signals appear in the region above each curve 
but are suppressed below it.

\begin{figure}
\begin{center}
\centerline{\psfig{figure=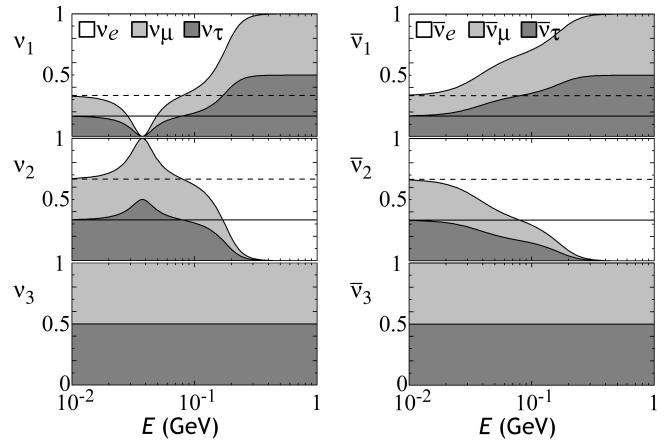,
width=\hsize}}
\caption{ \label{nuflav}
Flavor content of the three neutrino eigenstates of $\h$ (left)
and the three antineutrino eigenstates of $\hb$ (right) 
as a function of energy. 
For the puma model,
the left-hand panel shows the energy dependences of 
$|U_{a'e}|^2$ (white),
$|U_{a'\mu}|^2$ (light grey),
and $|U_{a'\ta}|^2$ (dark grey)
for each neutrino mass eigenstate $\nu_{a'}$,
$a' = 1,2,3$,
while the right-hand panel displays
the analogous energy dependences for antineutrinos.
For the \msm,
the corresponding quantities 
$|U_{a'e}|^2$ (regions above dashed lines),
$|U_{a'\mu}|^2$ (regions between dashed and solid lines),
and $|U_{a'\ta}|^2$ (regions below solid lines)
for neutrinos 
and those for antineutrinos are energy independent.
The models coincide at all energies 
for the eigenstates $\nu_3$, $\ol\nu_3$,
but $\nu_1$, $\ol\nu_1$ 
match $\nu_2$, $\ol\nu_2$ only at low energies.
}
\end{center}
\end{figure}

Figure \ref{kmplot} shows this plot
for the puma model with values \rf{c8a5m}
and the \msm.
The \msm\ has two independent oscillation lengths,
$L_\odot=4\pi E/\De m^2_\odot$ 
and $L_\text{atm}=4\pi E/\De m^2_\text{atm}$,
both of which grow linearly with the energy
and are therefore represented by straight lines in the plot.
In the puma model,
however,
the unconventional energy dependences
from $B(E)$ and $C(E)$ 
produce more general curves instead.
These curves partially differ for neutrinos and antineutrinos,
a consequence of the CPT violation
implied by the values \rf{c8a5m}.

The figure shows that
the puma curves merge with 
the \msm\ lines $L_\odot$ and $L_\text{atm}$ 
at low and high energies, 
respectively,
suggesting consistency of the puma model 
with results in KamLAND, 
solar, 
and atmospheric experiments.
This agreement is confirmed in the subsections below.
However,
the two models are qualitatively different 
at intermediate energies.

Novel effects arise from 
the unconventional energy dependence of $\h$,
which generates energy-dependent mixing.
The flavor content of the three eigenstates of $\h$ 
therefore changes with energy.
Figure \ref{nuflav}
shows this energy dependence
for the values \rf{c8a5m}. 
At low energies,
the flavor content approaches the tribimaximal limit. 
However,
at high energies the eigenstate $\nu_2$ 
becomes completely populated by $\nu_e$.
This implies the mixing $\nu_\mu\mix\nu_\ta$ is maximal 
and controlled by $\De_{31}$. 
The onset of this feature coincides
with the onset of the Lorentz-violating seesaw. 
Indeed,
as the mass term $A$ becomes negligible in $\h$, 
the fraction of $\nu_e$ in $\nu_2$ 
grows with the separation between the lines 
$\ltt$ and $\lto$ in Fig.\ \ref{kmplot}.

Notice that the mixing angles in the \msm\ 
are energy independent parameters
that can freely be chosen to match data.
In contrast,
the mixing angles in the puma model
at low and high energies
are determined by the texture of $\h$
and therefore are fixed features of the model
that cannot be adjusted according to experiment. 
This reduced freedom is one reason why 
the puma model offers a more economical description
of confirmed neutrino data than the \msm.

\begin{figure}
\begin{center}
\centerline{\psfig{figure=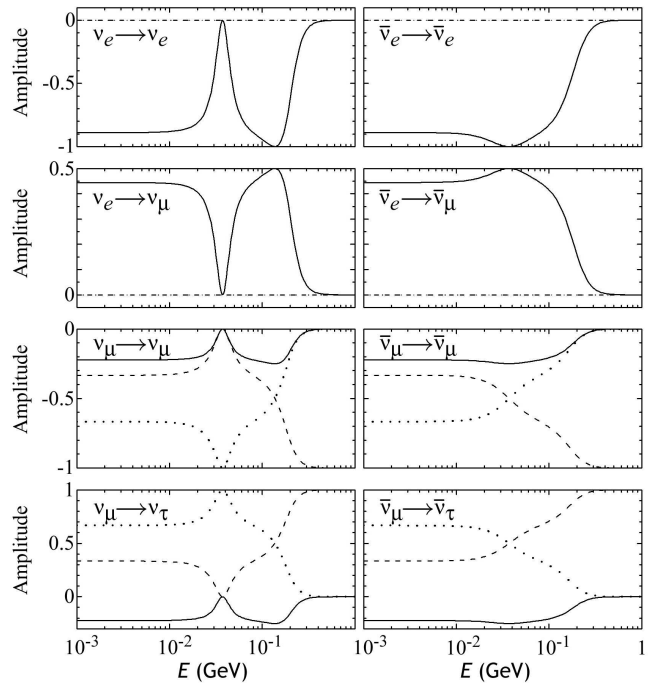,
width=\hsize}}
\caption{ 
Energy dependence of the oscillation amplitudes in the puma model.
In each flavor channel,
the amplitude factors $-4UUUU$ in Eq.\ \rf{probs} 
are plotted for each of the three $(a',b')$ values 
21 (solid lines),
31 (dashed lines),
and 32 (dotted lines).}
\label{amplitudesfig}
\end{center}
\end{figure}

The energy dependence of the mixing matrix $U$ 
implies the oscillation amplitudes $-4UUUU$ 
in each flavor channel 
and the corresponding probability \rf{probs}
are also energy dependent.
For given flavors $a$, $b$,
the oscillation amplitudes are shown in Fig.\ \ref{amplitudesfig}.
Note that negative amplitudes occur
for disappearance channels,
while positive amplitudes occur for appearance channels. 
The $S_2$ symmetry of $\h$ 
implies the four amplitudes for
$\nu_e\to\nu_\ta$,
$\nub_e\to\nub_\ta$,
$\nu_\ta\to\nu_\ta$,
and $\nub_\ta\to\nub_\ta$
are identical to those shown 
in the corresponding four central panels 
in the figure.
The low-energy $S_3$ symmetry of $\h$ 
forces the low-energy amplitudes to values
set by tribimaximal mixing
and ensures the low-energy equalities
$P_{\nu_e\to\nu_e}= P_{\nu_\mu\to\nu_\mu}= P_{\nu_\ta\to\nu_\ta}$
and 
$P_{\nub_e\to\nub_e}= 
P_{\nub_\mu\to\nub_\mu}= P_{\nub_\ta\to\nub_\ta}$. 
At high energies,
the amplitudes become either zero or one
due to the Lorentz-violating seesaw mechanism. 
The lower four panels in the figure
reveal that the dominant amplitude at high energies
has $(a',b')=(3,1)$,
leading to maximal $\nu_\mu\mix\nu_\ta$ mixing
and to an oscillation phase proportional 
to $\De_{31}$ and hence to $1/E$.
Note also that 
the zero component $U_{3e}$ of the mixing matrix \rf{U},
which is a consequence of the null eigenvalue of $\h$,
implies that the oscillation in any channel 
involving $\nu_e$ or $\nub_e$
is controlled by only one amplitude
because the other two vanish.
 
The three figures reveal many of the evolving properties 
associated with $\h$ and $\hb$ at intermediate energies.
For example,
a peak appears between 10 MeV and 100 MeV
in the $\lto$ curve for neutrinos in Fig.\ \ref{kmplot},
accompanied by corresponding features 
in Figs.\ \ref{nuflav} and \ref{amplitudesfig}.
The peak represents a divergence in $\lto$,
which occurs when $\De_{a'b'}$ vanishes.
Using the exact expressions \rf{las(puma)}
for the eigenvalues of $\h$,
we find that in general peaks occur
for all positive energies $E$ solving the equation 
\beq
A(B-C) + B^2 = 0.
\label{peaks}
\eeq
The peaks can in general occur 
for both neutrinos and antineutrinos.
The absence of these features 
in the antineutrino plots 
suggests an origin in CPT violation. 
Since coefficients for CPT-odd Lorentz violation 
reverse sign under a CPT transformation, 
the nature of the solutions to Eq.\ \rf{peaks}
for antineutrinos changes.
For the values \rf{c8a5m}, 
a single positive energy solves this equation for neutrinos,
but no solutions exist for antineutrinos
and hence no antineutrino peaks arise in Fig.\ \ref{kmplot}.

\subsection{Reactor antineutrinos}

In the puma model, 
the general survival probability for reactor antineutrinos is 
\beq
P_{\nub_e\to\nub_e} = 
1 - 16\fr{(A+\ol B)^4}{\ol N_1^2\ol N_2^2}
\sin^2\left(\half \De_{21}L \right).
\label{Pee_exact}
\eeq
At low energies,
the $A$ term in $\hb$ dominates.
Using the low-energy limits
$\ol N_1^2 \to 6A^2$,
$\ol N_2^2 \to 3A^2$,
we find for $P_{\nub_e\to\nub_e}$  
the simple low-energy approximation
\beq
P_{\nub_e\to\nub_e} \approx 
1 - \frac 89 \sin^2 \left(\fr{3m^2L}{4E} \right)
\quad
{(\rm low~energy)}.
\label{Pee(puma)}
\eeq
The fixed value 8/9 for the oscillation amplitude 
matches expectations because
at low energies $\hb$ is diagonalized 
using the tribimaximal mixing matrix.
This result applies to reactor antineutrinos 
in both long- and short-baseline experiments.
The large disappearance amplitude for reactor antineutrinos 
is evident in the $\nub_e\to\nub_e$ panel
of Fig.\ \ref{amplitudesfig}.

\begin{figure}
\begin{center}
\centerline{\psfig{figure=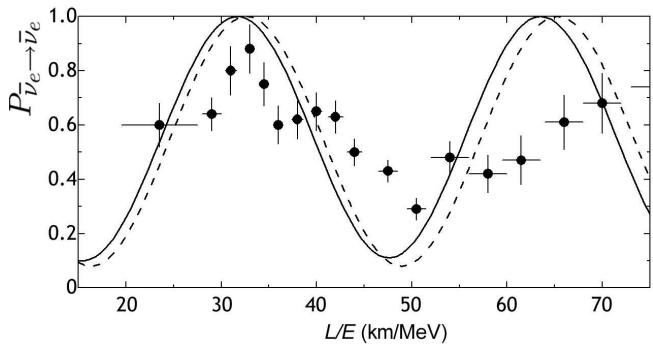,
width=\hsize}}
\caption{ \label{kamland}
Reactor-antineutrino survival probabilities 
as a function of $L/E$ in the puma model (solid line) 
and in the \msm\ (dashed line).
The data are from the long-baseline KamLAND experiment,
for which $L\simeq180$ km 
\cite{KamLAND(L/E)}.}
\end{center}
\end{figure}

\subsubsection{Long-baseline reactor: KamLAND}
\label{Sec: LB_reactor}

In the \msm, 
the reactor-antineutrino survival probability 
for long-baseline experiments is
\beq
P^{3\nu{\rm SM}}_{\nub_e\to\nub_e} 
\approx
1 - \sin^22\th_{12}
\sin^2{\left(\fr{\De m^2_\odot L}{4E}\right)}.
\label{Peemsmlong}
\eeq
The data indicate values for the \msm\ parameters of 
$\sin^22\th_{12}\simeq0.92$
and $\De m^2_\odot\simeq7.58\times10^{-5}$ eV$^2$ 
\cite{KamLAND}. 
Comparing the oscillation phase in this result 
with that in Eq.\ \rf{Pee(puma)},
we find that agreement with the KamLAND results 
can be achieved by choosing
the mass parameter $m^2$ to be $m^2=\De m^2_\odot/3$ 
\cite{puma}.
This gives the numerical value adopted in Eq.\ \rf{c8a5m}. 
The match between the two models
is shown in Fig.\ \ref{kamland}.

Notice that the disappearance of reactor antineutrinos 
is described using only one parameter $m$
instead of the usual two in the \msm. 
The conventional solar mixing angle $\th_{12}$ 
is eliminated as a degree of freedom 
by the form of the texture $\h$.
Inspecting Eq.\ \rf{Pee(puma)}
reveals that at low energies
the effective value of $\sin^22\th_{12}$ 
is numerically fixed to 
$(\sin^22\th_{12})_{\rm eff}\simeq 0.89$,
which is close to the measured magnitude.
The reader is however cautioned
that this interpretation fails at higher energies 
due to the energy dependence of the mixing in the puma model. 

\subsubsection{Short-baseline reactors}
\label{Short-baseline reactors}

In recent years,
numerous reactor experiments with short baselines
$L\lsim 1$ km
such as 
Bugey ($L\simeq 15,40$ m) 
\cite{Bugey}, 
CHOOZ ($L\simeq 1$ km) 
\cite{CHOOZ}, 
G\"osgen ($L\simeq 38,46,65$ m) 
\cite{Gosgen}, 
and Palo Verde ($L\simeq 750, 890$ m) 
\cite{Palo Verde} 
have sought evidence for
the disappearance of electron antineutrinos 
with null results. 
The explanations of these results
differ qualitatively 
in the puma model and the \msm.

In the \msm, 
the $\ol\nu_e$ survival probability is
\beq
P^{3\nu{\rm SM}}_{\nub_e\to\nub_e} 
\approx 1 - \sin^22\th_{13}
\sin^2{\left(\fr{\De m^2_{\rm atm}L}{4E}\right)}.
\label{Pee(3nuSM)SB}
\eeq
For energies $E\simeq3$ MeV,
this gives an antineutrino disappearance length
$2\pi E/\De m^2_{\rm atm} \simeq
8.1\times 10^{18}$ GeV$^{-1}$,
which is about 1.5 km.
The null experimental results
are therefore interpreted in the \msm\
as a consequence of a small mixing angle $\th_{13}$.
Note that the \msm\ survival probabilities 
\rf{Peemsmlong} and \rf{Pee(3nuSM)SB} 
for long and short baselines,
respectively,
have the same form but involve four different parameters,
$\De m^2_{\rm solar}$,
$\th_{12}$,
$\De m^2_{\rm atm}$, 
and $\th_{13}$.

In contrast,
in the puma model
the oscillation probability \rf{Pee(puma)}
holds at low energies for any baseline.
Only the single parameter $m$ is required to describe 
both the long- and short-baseline data.
For energies $E\simeq3$ MeV,
the antineutrino disappearance length is
$\lttb\equiv\pi/\ol\De_{21}
\approx 2\pi E/\De m^2_\odot
\simeq 2.5\times 10^{20}$ GeV$^{-1}$,
which is about 50 km.
The null reactor results 
are therefore understood in this model 
as a consequence of the short baselines,
which limit the contribution of the oscillation phase
to the survival probability,
rather than a consequence of a small oscillation amplitude 
as in the \msm.
Indeed,
the amplitude of the oscillating term 
in Eq.\ \rf{Pee(puma)} is 8/9, 
which is large.

Since the puma model contains no term
with a phase involving $\De m^2_{\rm atm}$,
we see that at low energies the effective value 
$(\sin^22\th_{13})_{\rm eff}$ 
of the \msm\ quantity $\sin^22\th_{13}$ 
is exactly zero.
This is a consequence of the zero value of $U_{3e}$,
as can be confirmed by comparing
the \msm\ mixing matrix
with the tribimaximal limit of the mixing matrix \rf{U}.
Note,
however,
that the energy dependence of the mixing matrix 
makes this result invalid at higher energies,
where the effective value 
$(\sin^22\th_{13})_{\rm eff}$ 
extracted from high-energy experiments
can be nonzero even though $U_{3e}$ identically vanishes.

\subsection{Solar neutrinos}

\begin{figure}
\begin{center}
\centerline{\psfig{figure=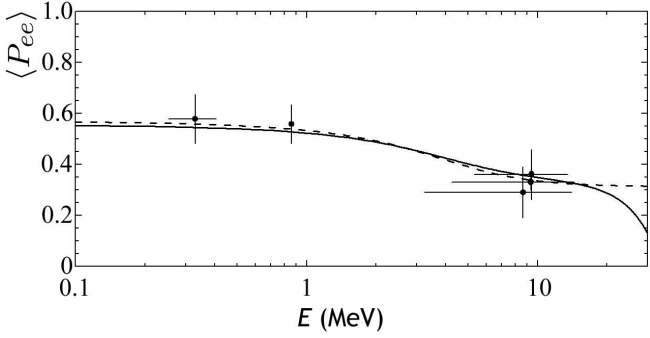,
width=\hsize}}
\caption{ 
Averaged survival probability for solar neutrinos 
in the puma model (solid line) 
and in the \msm\ (dashed line). 
Both cases include matter-induced effects 
in the adiabatic approximation.
The data are from Ref.\ \cite{Borexino}.}
\label{solarfig}
\end{center}
\end{figure}

For neutrinos propagating in matter,
the effective hamiltonian $\h$
acquires an additional term
\cite{msw}.
The modified effective hamiltonian $(\h)^{\rm M}$
in the solar interior can be written as 
\beq
(\h)_{ab}^{\rm M}=
(\h)_{ab}+V_\odot\de_{ae}\de_{be},
\label{h(solar)}
\eeq
where the solar matter potential $V_\odot$
takes the value
$V_\odot =\sqrt2 G_{\rm F}n_e 
\simeq 7.84\times10^{-21}$ GeV 
at the solar core 
\cite{Bahcall_2006}. 

The presence of the solar potential 
preserves the puma texture \rf{h(puma)}
because it corresponds to a simple redefinition 
of the function $C$ 
of the form $C \to C+V_\odot$.
The exact eigenvalues and the exact mixing matrix 
in the presence of matter 
can therefore be found immediately 
by applying this redefinition to 
Eqs.\ \rf{las(puma)}, \rf{U}, and \rf{normalization}. 
A short calculation reveals that
the averaged survival probability 
of solar neutrinos takes the exact form 
\bea
\langle P_{\nu_e\to\nu_e} \rangle &=& 
\left(\fr{(\la_1^{\rm M}-2A)}{N_1^{\rm M}}
\fr{(\la_1-2A)}{N_1}\right)^2
\nn\\
&&
+ \left(\fr{(\la_2^{\rm M}-2A)}{N_2^{\rm M}}
\fr{(\la_2-2A)}{N_2}\right)^2.
\label{PeeSolar(puma)}
\eea

For the lower-energy region of the solar spectrum 
with $E\sim 0.1$ MeV, 
the solar potential $V_\odot$
and the functions $B$ and $C$ are negligible.
In this limit,
the averaged survival probability becomes
\beq
\langle P_{\nu_e\to\nu_e} \rangle
\approx \sum_{a'} |U_{a'e}|^4
=\frac{5}{9}
\quad
{(\rm low~energy)},
\eeq
in agreement with the data. 
This result is to be expected 
because the vacuum mixing matrix 
is tribimaximal at low energies. 

For higher energies, 
the solar potential and the Lorentz-violating terms 
can introduce novel effects,
depending on the form of the functions $B$ and $C$.
The detailed form of the averaged survival probability
therefore becomes model dependent.
However,
the neutrino survival probability 
initially drops below the limiting value 5/9
as the energy increases.
This generic effect is a consequence 
of the energy independence of $V_\odot$,
which ensures $V_\odot$ becomes relevant
at energies comparable or below those 
for the Lorentz-violating terms
and thereby enhances the disappearance of $\nu_e$.

The above features are visible 
in Fig.\ \ref{solarfig}.
The solid line displays the averaged survival probability
for the values \rf{c8a5m}. 
The curve is similar to that obtained from the \msm\
and is compatible with observations.

\subsection{Atmospheric neutrinos}
\label{Sec: atm}

\begin{figure}
\begin{center}
\centerline{\psfig{figure=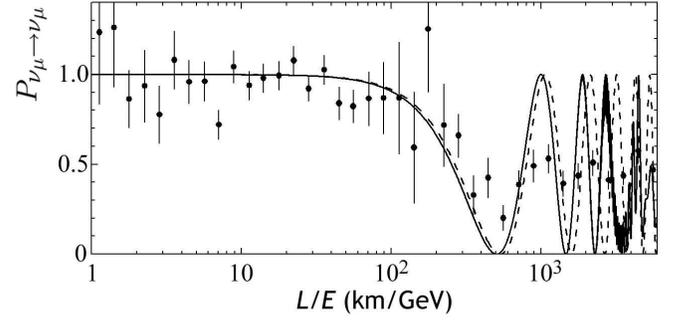,
width=\hsize}}
\caption{ 
Survival probability for atmospheric neutrinos 
as a function of $L/E$ 
in the puma model (solid line) and in the \msm\ (dashed line)
compared to SK data,
for which $L_{\rm SK}\simeq600$ km 
\cite{SK(L/E)}.}
\label{Fig: SK_c8a5m1}
\end{center}
\end{figure}

In the puma model, 
the exact survival probability of atmospheric neutrinos is
\bea
P_{\nu_\mu\to\nu_\mu} &=& 
1 - 4\fr{(A+B)^4}{N_1^2N_2^2}
\sin^2\left(\half \De_{21}L \right)
\nn\\&&
- 2\fr{(A+B)^2}{N_1^2}\,\sin^2\left(\half \De_{31}L \right)
\nn\\&&
- 2\fr{(A+B)^2}{N_2^2}\,\sin^2\left(\half \De_{32}L \right). 
\label{Pmm_exact}
\eea
However,
as $E$ grows the $A$ term becomes negligible, 
so the sole mass parameter $m$
is irrelevant for high-energy oscillations.
Requiring $C$ to increase with energy faster than $B$ 
yields the high-energy limits
$\la_1 \to -{2B^2}/{C}$,
$\la_2 \to C$, 
$N_1^2 \to 2B^2$, 
and $N_2^2 \to C^2$. 
The electron-neutrino content then lies exclusively 
in the second eigenstate.
This leaves the other two uniformly populated 
by $\nu_\mu$ and $\nu_\ta$, 
as can be verified by examining Fig.\ \ref{nuflav}.
The survival probability \rf{Pmm_exact} 
then takes the simple form
\bea
P_{\nu_\mu\to\nu_\mu} &\approx&
1 - \sin^2\left(\fr{B^2L}{C}\right)
\quad
{(\rm high~energy)}
\nn\\
&\approx&
1 - \sin^2\left(\fr{(\k{p})^2L}{\C{q}E}\right), 
\label{Pmm(puma)}
\eea
where in the second equation 
the leading contributions to $B$ and $C$
are expressed as monomials of the form \rf{B,C(E)}.
Note that the unit amplitude of the oscillation term 
implies maximal mixing,
as discussed following Eq.\ \rf{pqlim}.

In contrast,
the \msm\ survival probability for atmospheric neutrinos 
takes the form 
\beq
P^{3\nu{\rm SM}}_{\nu_\mu\to\nu_\mu} \simeq 
1 - \sin^22\th_{23}
\sin^2{\left(\fr{\De m^2_{\rm atm}L}{4E}\right)}
\label{3nusmatmos}
\eeq
depending on two parameters $\th_{23}$ and $\De m^2_{\rm atm}$.
Experimental data provide the values
$\sin^22\th_{23}>0.90$ 
and $|\De m^2_{\rm atm}|\simeq2.32\times10^{-3}$ eV$^2$ 
\cite{MINOSnu2011}.
Comparison of Eqs.\ \rf{Pmm(puma)} and \rf{3nusmatmos}
suggests agreement with atmospheric data can be obtained 
when the ratio of $(\k{p})^2$ and $\C{q}$ satisfies
\beq
\fr{(\k{p})^2}{\C{q}} = \frac 14 \De m^2_{\rm atm}.
\label{match_atm}
\eeq
This condition has been used
to constrain the coefficients $\a{5}$ and $\C{8}$ 
in Eq.\ \rf{c8a5m} 
\cite{puma}.
The resulting match between the two models
is shown in Fig.\ \ref{Fig: SK_c8a5m1}
along with SK data.

Note that the ratio \rf{match_atm} 
represents only one degree of freedom.
Nonetheless,
it suffices to reproduce the data 
for atmospheric neutrinos
via Eq.\ \rf{Pmm(puma)}.
The other degree of freedom in the two coefficients 
$\k{p}$, $\C{q}$ 
determines the onset of the Lorentz-violating seesaw.
Increasing $\C{q}$ while holding fixed the ratio \rf{match_atm}
causes the seesaw to trigger at lower energies.

\subsection{Short-baseline accelerator neutrinos}
\label{Sec: SB_acc}

At high energies $E\gsim 1$ GeV, 
a variety of short-baseline experiments 
have reported null results.
BNL-E776 ($L=1$ km) 
searched for $\nu_\mu\to\nu_e$
and $\ol\nu_\mu\to\ol\nu_e$ 
at 1 GeV
\cite{BNL}.
CCFR ($L\simeq 1$ km) 
searched for $\nu_\mu\to\nu_e$, 
$\ol\nu_\mu\to\ol\nu_e$, 
$\nu_e\to\nu_\ta$, 
and $\ol\nu_e\to\ol\nu_\ta$ 
at 140 GeV
\cite{CCFR}.
CDHS ($L\simeq 130$ m) 
searched for $\nu_\mu$ disappearance 
at 1 GeV
\cite{CDHS}.
CHORUS ($L\simeq 600$ m) 
searched for $\nu_\mu\to\nu_\ta$ 
at 27 GeV 
\cite{CHORUS}.
NOMAD ($L\simeq 600$ m) 
searched for $\nu_\mu\to\nu_\ta$ 
and $\nu_e\to\nu_\ta$ 
at 45 GeV 
\cite{NOMAD}. 
NuTeV ($L\simeq 1$ km) 
searched for $\nu_\mu\to\nu_e$
and $\ol\nu_\mu\to\ol\nu_e$ 
at 150 GeV
\cite{NuTeV}. 

The puma model is consistent with 
all these null results.
For energies above the seesaw scale $\sim 1$ GeV,
$\nu_\mu\mix\nu_\ta$ 
mixing becomes maximal by construction,
as described following Eq.\ \rf{pqlim}.
This feature implies vanishing high-energy mixing 
and hence no oscillations in the channels 
$\nu_\mu\to\nu_e$, 
$\ol\nu_\mu\to\ol\nu_e$, 
$\nu_e\to\nu_\ta$, 
and $\ol\nu_e\to\ol\nu_\ta$.
The behavior can be seen directly from 
Fig.\ \ref{amplitudesfig},
which displays the energy dependence 
of the oscillation amplitudes.

In the $\nu_\mu\to\nu_\ta$ channel,
the oscillation amplitude is maximal at high energies.
However,
the oscillation phase is controlled by $\De_{21}$,
which generates an appearance length $\ltt$ 
of several hundred kilometers at 1 GeV. 
The lack of a signal in this channel 
in the CHORUS or NOMAD data
is therefore understood here
as a consequence of their short baselines.

\subsection{MiniBooNE anomalies}
\label{Sec: MB_anomaly}

\begin{figure}
\begin{center}
\centerline{\psfig{figure=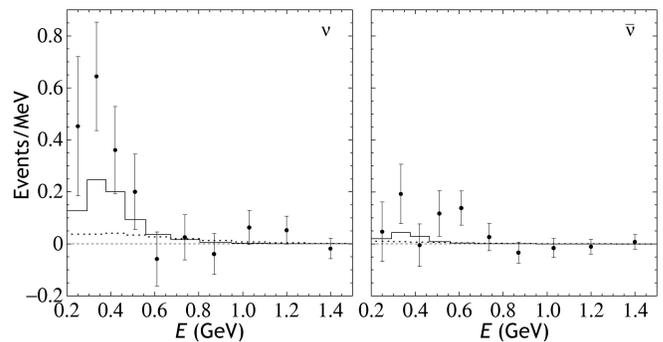,
width=\hsize}}
\caption{ 
Comparison of the puma model 
(solid lines; 
$\ch^2_\nu=1.0$,
$\ch^2_{\ol\nu}=0.9$),
the tandem model 
\cite{tandem} 
(dotted lines;
$\ch^2_\nu=1.9$,
$\ch^2_{\ol\nu}=1.0$),
and the \msm\ 
(dashed lines; 
$\ch^2_\nu=2.2$,
$\ch^2_{\ol\nu}=1.1$),
with MiniBooNE neutrino \cite{MiniBooNE1}
and antineutrino \cite{MiniBooNE2} data.
}
\label{MBfig}
\end{center}
\end{figure}

Two results from the MiniBooNE experiment 
indicate possible oscillation effects that
cannot be accommodated within the \msm.
For neutrino oscillations 
$\nu_\mu\to\nu_e$,
MiniBooNE finds a 3$\si$ excess of events 
at low energies around 200-500 MeV 
\cite{MiniBooNE1}. 
For antineutrino oscillations
$\ol\nu_\mu\to\ol\nu_e$,
a 1.3$\si$ low-energy excess has also been reported 
\cite{MiniBooNE2},
with recent preliminary data suggesting a larger excess
\cite{edz}. 

These results are interesting in the present context
because they lie in the energy region 
where the seesaw mechanism is triggered. 
Following onset of the seesaw,
the eigenvalue $\la_1$ decreases linearly with energy
while $\la_2$ grows rapidly. 
The appearance length $\ltt\propto(\la_2-\la_1)^{-1}$ 
therefore drops steeply,
becoming a few hundred meters at MiniBooNE energies.
This produces a large oscillation phase 
and hence a signal in the experiment.
However,
the oscillation amplitude for $\nu_\mu\mix\nu_e$ mixing 
rapidly goes to zero as the $\nu_\mu\mix\nu_\ta$ mixing 
becomes maximal, 
as can be seen in Fig.\ \ref{amplitudesfig}.
As a result, 
the appearance signal in MiniBooNE vanishes at higher energies. 

The puma model therefore naturally describes 
a low-energy excess in MiniBooNE. 
Moreover, 
the excess can differ substantially for neutrinos and antineutrinos 
when a coefficient for CPT-odd Lorentz violation is involved,
such as occurs for the \ceafm\ example \rf{c8a5m}.
In general,
the energy at which the excess appears 
depends on the seesaw scale
and becomes smaller as $\C{q}$ increases.
For the values \rf{c8a5m}, 
the match to data is shown in Fig.\ \ref{MBfig}.

We emphasize that these interesting features 
of the model arise without introducing 
additional particles or forces.
They are a consequence of the comparatively elegant texture
\rf{h(puma)}
that describes all compelling neutrino-oscillation data.

\section{Predictions}
\label{Predictions}

The discussion in the previous sections demonstrates that
two of the three parameters of the model \rf{h(puma)} 
suffice to reproduce all the compelling data 
for neutrino and antineutrino oscillations,
while the third accommodates the two MiniBooNE anomalies.
Comparison to the \msm,
which uses five nonzero parameters
to describe established results 
but cannot reproduce the MiniBooNE anomalies,
suggests the puma model offers 
a frugal interpretation of known data.

The model predicts a variety of signals,
some of which differ qualitatively from \msm\ expectations. 
In this section,
we address some features of relevance to future experiments.

\begin{figure}
\begin{center}
\centerline{\psfig{figure=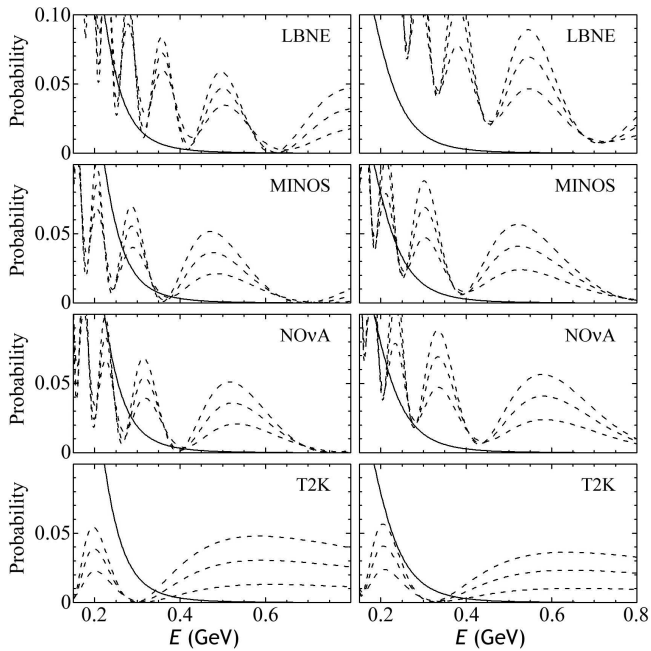,
width=\hsize}}
\caption{ 
Prediction for the probabilities 
of $\nu_e$ appearance (left) 
and of $\nub_e$ appearance (right) 
in various long-baseline experiments 
according to the puma model (solid lines) 
and the \msm\ 
(upper dashed lines, $\sin^22\th_{13}=0.02$;
middle dashed lines, $\sin^22\th_{13}=0.05$; 
lower dashed lines, $\sin^22\th_{13}=0.08$). 
Matter effects are included.
}
\label{lbnuefig}
\end{center}
\end{figure}

\begin{figure}
\begin{center}
\centerline{\psfig{figure=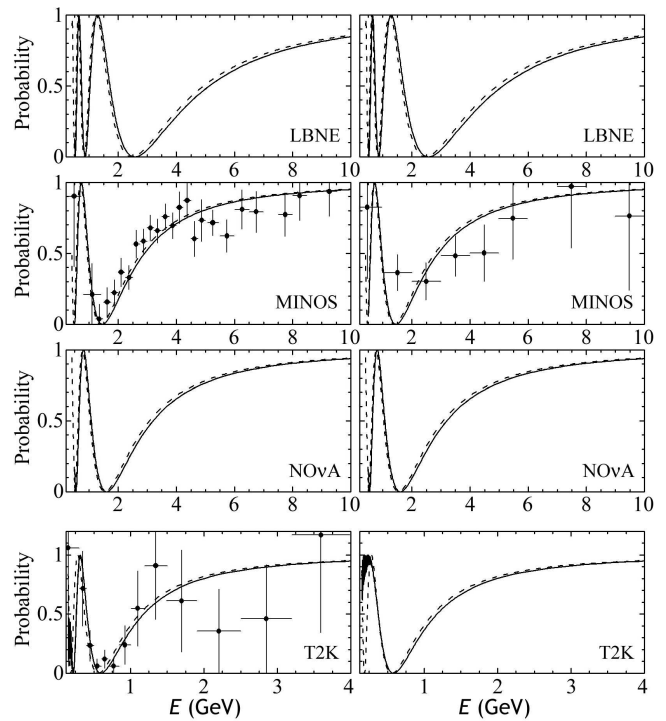,
width=\hsize}}
\caption{ 
Prediction for the probabilities
of $\nu_\mu$ disappearance (left) 
and of $\nub_\mu$ disappearance (right) 
in various long-baseline experiments 
according to the puma model (solid lines) 
and the \msm\ (dashed lines).
Matter effects are included.
The data are taken 
from Refs.\ \cite{MINOSanomaly,t2kdata,MINOSnu2011}.
} 
\label{lbnumfig}
\end{center}
\end{figure}

\subsection{Long-baseline neutrinos}

A variety of long-baseline experiments,
including LBNE ($L\simeq 1300$ km) 
\cite{LBNE}, 
MINOS ($L\simeq 735$ km) 
\cite{MINOSnu_e}, 
NO$\nu$A ($L\simeq 810$ km) 
\cite{NOvA}, 
and T2K ($L\simeq 298$ km) 
\cite{T2K}, 
have design capabilites to search 
for $\nu_e$ appearance in a $\nu_\mu$ beam. 
These searches are motivated in part
by the prospects of measuring the \msm\ parameter $\th_{13}$.
No such parameter exists in the model \rf{h(puma)},
but signals in these experiments may nonetheless appear. 

To characterize potential signals in the puma model,
recall that the appearance length 
$\ltt$ for $\nu_\mu\to\nu_e$ decreases steeply with energy 
due to the seesaw mechanism. 
Baselines $L\gg \ltt$ therefore involve rapid oscillations, 
so accelerator experiments with long baselines 
can observe only the averaged oscillation probability,
given exactly by
\beq
\langle P_{\nu_\mu\to\nu_e} \rangle = 
4\fr{(A+B)^4}{(N^M_1 N_2^M)^2}.
\label{PemAve(puma)}
\eeq
To allow for matter effects on neutrinos traversing the Earth,
$N_1^M$ and $N_2^M$ 
are given by Eq.\ \rf{normalization} 
with the replacement $C \to C+V_\oplus$,
where the Earth's matter potential $V_\oplus$
is $V_\oplus\simeq 1.2\times10^{-22}$ GeV. 

For energies above the seesaw scale,
$\nu_\mu\mix\nu_\ta$ mixing dominates
while $\nu_\mu\mix\nu_e$ mixing is highly suppressed.
However,
to describe the SK and MINOS data, 
the seesaw must trigger below 1 GeV.
This means only small signals from $\nu_\mu\to\nu_e$ transitions 
can appear in the high-energy experiments
LBNE, MINOS, and NO$\nu$A.  
In contrast,
T2K runs at lower energies,
and so a larger appearance signal that decreases 
rapidly with the energy is to be expected. 
Quantitative predictions 
for the probabilities for $\nu_e$ and $\ol\nu_e$ apppearance 
in the various experiments 
are shown in Fig.\ \ref{lbnuefig}
for the values \rf{c8a5m}.
Note that matter effects are almost negligible
compared to the large eigenvalue $\la_2$ controlling the mixing,
whereas for the \msm\ curves 
they induce substantial differences 
between the probabilities for $\nu_e$ and $\ol\nu_e$ appearance. 

We remark in passing that attempting to interpret 
these signals as arising from
a nonzero \msm\ angle $\th_{13}$
would predict that $\th_{13}$ is larger in T2K
than in the other higher-energy experiments.
This is compatible with recent
results for $\nu_e$ appearance
\cite{t2krecent}. 
Note also that within this perspective
the effective values of $\th_{13}$
obtained with long-baseline accelerators 
are unrelated to the effective values of $\th_{13}$
extracted from studies of reactor antineutrinos
discussed in Sec.\ \ref{Short-baseline reactors}.

High-energy long-baseline experiments can also perform 
precision studies of $\nu_\mu$ disappearance. 
For this oscillation channel, 
no differences between the puma model and the \msm\
are expected in LBNE, MINOS, and NO$\nu$A 
because they operate at energies above the seesaw scale. 
However,
significant differences between neutrinos and antineutrinos 
are predicted for the lower-energy portion of the T2K spectrum
when the theory contains 
a coefficient for CPT-odd Lorentz violation,
as in the \ceafm\ model.
The predictions are displayed in Fig.\ \ref{lbnumfig},
where as before
the values \rf{c8a5m} are used for illustration.

\subsection{Short-baseline neutrinos}

\begin{figure}
\begin{center}
\centerline{\psfig{figure=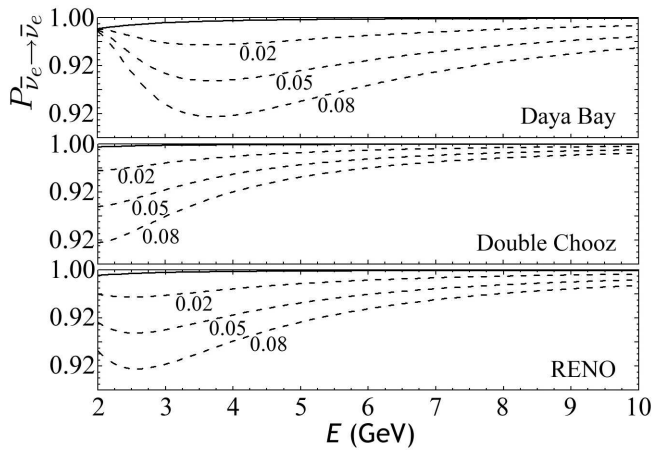,
width=\hsize}}
\caption{ 
Predictions for the probability of $\nub_e$ disappearance 
in the Daya Bay, Double Chooz, and RENO experiments
according to the puma model (solid lines) 
and the \msm\ (dashed lines,
labeled with the value of $\sin^22\th_{13}$). 
} 
\label{sbacfig}
\end{center}
\end{figure}

\begin{figure}
\begin{center}
\centerline{\psfig{figure=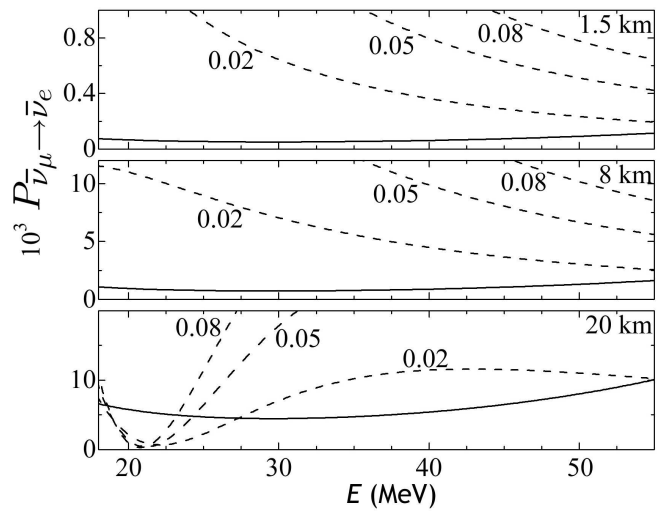,
width=\hsize}}
\caption{ 
Predictions for the probability of $\ol\nu_e$ appearance 
at the three baselines proposed for the DAE$\de$ALUS experiment
according to the puma model (solid lines) 
and the \msm\ (dashed lines,
labeled with the value of $\sin^22\th_{13}$). 
} 
\label{daedalusfig}
\end{center}
\end{figure}

Using baselines in the range 1-2 km, 
modern reactor experiments such as
Daya Bay ($L\simeq 1985$ m) 
\cite{DayaBay}, 
Double Chooz ($L\simeq 1050$ m) 
\cite{Double Chooz}, 
and RENO ($L\simeq 1380$ m) 
\cite{RENO}
propose to measure the disappearance of reactor antineutrinos. 
Like their long-baseline cousins,
these experiments are driven partly by prospects
for measuring the \msm\ mixing angle $\th_{13}$. 

In the puma model, 
the $\nub_e$ survival probability 
for energies 2-9 MeV 
has a large amplitude,
as can be seen from Eq.\ \rf{Pee(puma)}. 
The oscillation signal in a given reactor experiment
therefore depends only on the size of the baseline $L$
compared to the antineutrino disappearance length
$\lttb\simeq 50$ km
discussed in Sec.\ \ref{Short-baseline reactors}.
The baselines for the 
Daya Bay, Double Chooz, and RENO experiments
are all short compared to this,
with Daya Bay having the greater sensitivity 
to oscillation signals due to its longer baseline.
Since $\lttb$ grows linearly with the energy, 
any oscillation signal in these experiments
is expected to appear predominantly at low energies.
Using the model values \rf{c8a5m}, 
the predictions for the disappearance probabilities
in the three experiments
are shown in Fig.\ \ref{sbacfig}.

We note in passing that the recent suggestion
of an overestimation of antineutrino fluxes 
in short-baseline reactor experiments
\cite{ReactorAnomaly}
is difficult to reconcile 
with the three-parameter puma model.
Since effects at low energies are governed
by only one parameter $m$,
which is fixed by KamLAND data, 
no other oscillation length appears at reactor energies.
The existence of only one parameter is a consequence 
of the $S_3$ flavor symmetry, 
so a slight breaking of this symmetry at low energies
could accommodate an additional parameter
and hence a corresponding signal. 
This construction would introduce 
another degree of freedom
but requires no additional neutrinos. 
However, 
investigations along these lines
lie beyond the scope of the present work.

Another experiment of interest 
in the context of short-baseline neutrinos 
is the recent DAE$\de$ALUS proposal 
\cite{DAEdALUS}
to study CP violation,
which would generate neutrinos 
at several different baseline distances from a detector
using high-power accelerator modules 
to beam protons onto graphite sources.
A popular configuration would offer the capability
to search for $\nub_\mu\to\nub_e$ transitions 
using three baselines of about 1.5 km, 8 km, and 20 km. 
The large oscillation amplitude in this region
suggests appearance signals in the detector 
can be expected from the more distant sources.
The predicted appearance probabilities
for the three proposed baselines
are shown in Fig.\ \ref{daedalusfig}.

\section{Variant puma models}
\label{Variant puma models}

In the preceding sections,
the implications of the general texture \rf{h(puma)}
have been illustrated with the \ceafm\ model,
using the specific values \rf{c8a5m}. 
However,
other models can be constructed using $\h$
that successfully describe most or all
compelling neutrino data.
Some of these offer distinctive features
or intriguing possibilities
for describing experimental anomalies
beyond the MiniBooNE ones.
This section outlines some results 
for a few of these variant models.

\subsection{The \cfatm\ model}
\label{Sec:model_c4a3m}

\begin{figure}
\begin{center}
\centerline{\psfig{figure=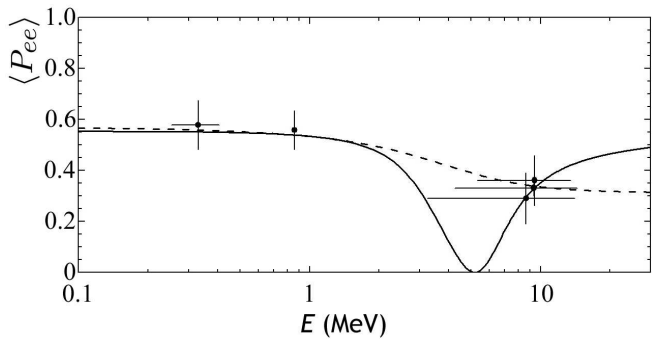,
width=\hsize}}
\caption{ 
Averaged survival probability for solar neutrinos 
in the \cfatm\ model (solid line) 
and in the \msm\ (dashed line). 
Both cases include matter-induced effects 
in the adiabatic approximation.
The data are taken from Ref.\ \cite{Borexino}.
} 
\label{solrenfig}
\end{center}
\end{figure}

\begin{figure}
\begin{center}
\centerline{\psfig{figure=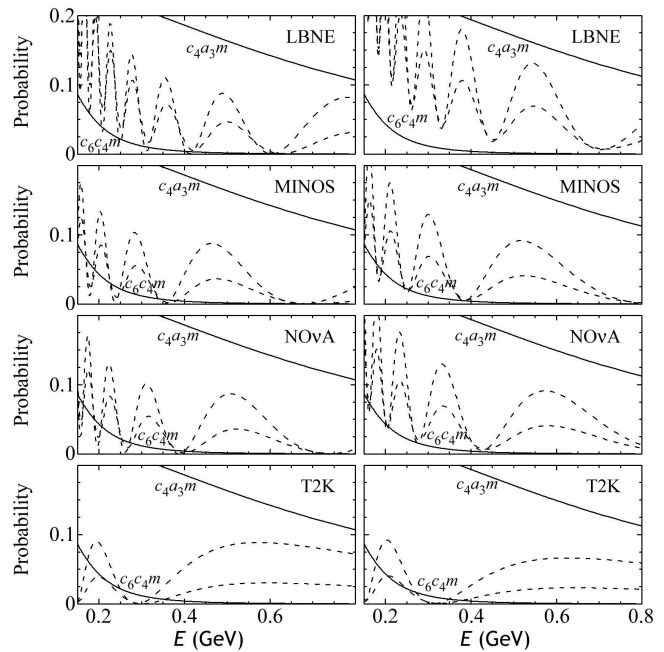,
width=\hsize}}
\caption{ 
Prediction for the probabilities 
of $\nu_e$ appearance (left) 
and of $\nub_e$ appearance (right) 
in various long-baseline experiments 
according to variant puma models (solid lines,
labeled by the model) 
and the \msm\ 
(upper dashed lines, $\sin^22\th_{13}=0.15$;
lower dashed lines, $\sin^22\th_{13}=0.05$). 
Matter effects are included.
} 
\label{pmerenfig}
\end{center}
\end{figure}

The texture $\h$ with the smallest monomial orders 
$p$ and $q$ in Eq.\ \rf{B,C(E)}
requires only renormalizable operators 
of dimensions 3 and 4 in the minimal SME,
hence producing a \cfatm\ model. 
For definiteness,
we adopt in this subsection the specific numerical values
\bea
m^2 &=& 2.6\times10^{-23} {\rm ~GeV}^2, 
\nn\\
\a{3} &=& -2.5\times10^{-21} {\rm ~GeV}, 
\nn\\
\C{4} &=& 1.0\times10^{-20}.
\label{c4a3m}
\eea
As mentioned at the end of Sec.\ \ref{Sec: puma_model},
we use a zero $ee$ entry in the $B$ term
for this model.
The coefficient $\a{3}$ comes with CPT violation,
so differences between neutrino and antineutrino properties
can be expected.
These values are consistent 
with limits from direct mass measurements,
cosmological mass bounds,
and constraints on anisotropic oscillations. 

The \cfatm\ model is compatible 
with all accepted experimental oscillation results
discussed in Sec.\ \ref{Experiments},
including those obtained with
reactor, solar, and atmospheric neutrinos.
However,
the eigenvalue $\la_2$ grows too slowly 
to produce a signal in MiniBooNE
because the function $C$ is linear in energy. 

An interesting qualitative difference 
introduced by the model 
appears in the predicted averaged survival probability
for solar neutrinos,
shown in Fig.\ \ref{solrenfig}. 
The probability curve incorporates 
a striking neutrino-disappearance maximum 
in the central-energy region,
despite passing though all data points.
This reflects the importance at lower energies 
of the coefficient $\a{3}$,
an effect absent for the $\a{5}$ coefficient
in the \ceafm\ model. 
The curve shape suggests future analyses of solar data 
in the 1-10 MeV part of the neutrino spectrum 
could provide an interesting experimental test of the model.

Another distinctive feature of the model
is a large signal for $\nu_\mu\to\nu_e$ oscillations
in long-baseline experiments.
The signal decreases with energy,
as shown in Fig.\ \ref{pmerenfig}. 
Analysis of the recent data 
supporting electron-neutrino appearance 
in the T2K and MINOS experiments
\cite{t2krecent}
could provide a sharp constraint on this signal, 
potentially excluding the values \rf{c4a3m}.

\subsection{The \cscfm\ model}

\begin{figure}
\begin{center}
\centerline{\psfig{figure=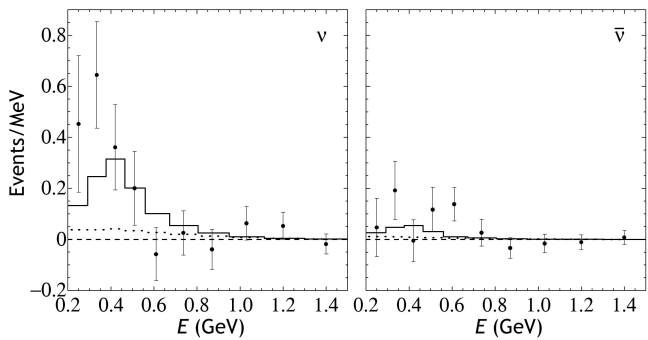,
width=\hsize}}
\caption{ 
Comparison of the \cscfm\ model 
(solid lines; 
$\ch^2_\nu=1.0$,
$\ch^2_{\ol\nu}=0.8$),
the tandem model 
\cite{tandem} 
(dotted lines;
$\ch^2_\nu=1.9$,
$\ch^2_{\ol\nu}=1.0$),
and the \msm\ 
(dashed lines; 
$\ch^2_\nu=2.2$,
$\ch^2_{\ol\nu}=1.1$),
with MiniBooNE neutrino \cite{MiniBooNE1}
and antineutrino \cite{MiniBooNE2} data.
}
\label{MBvariantfig}
\end{center}
\end{figure}

Other interesting variant models 
can be constructed using only CPT-even operators. 
Using monomials of smallest order produces a \cscfm\ model. 
We choose here the specific numerical values
\bea
m^2 &=& 2.6\times10^{-23} {\rm ~GeV}^2, 
\nn\\
\C{4} &=& 7.7\times10^{-20}, 
\nn\\
\C{6} &=& 1.0\times10^{-17} {\rm ~GeV}^{-2}.
\label{c6c4m}
\eea
As for other examples considered in this work,
these values are consistent 
with limits from direct mass measurements,
cosmological mass bounds,
and constraints on anisotropic oscillations. 

Like the \ceafm\ and \cfatm\ models,
the \cscfm\ model provides a good match 
to the data from reactor, solar, and atmospheric neutrinos
discussed in Sec.\ \ref{Experiments}.
The presence of the $\C{4}$ term substantially affects the physics
in the region 10 MeV to 1 GeV. 
It generates a signal in the MiniBooNE region
that includes low-energy excesses
in both neutrinos and antineutrinos,
as shown in Fig.\ \ref{MBvariantfig}. 
Note the asymmetry in the signal,
which here reflects experimental acceptance
rather than CPT violation.

For $\nu_e$ appearance in long-baseline experiments, 
the probabilities are generically closer in magnitude
to those of the \msm\ with a moderate value of $\th_{13}$,
as shown in Fig.\ \ref{pmerenfig}. 
One interesting feature
is the substantially larger signal produced in T2K
relative to MINOS,
which is compatible with the central values
of recently reported measurements
\cite{t2krecent}.

\subsection{Four-coefficient models}
\label{Sec:model_c8a5c4m}

\begin{figure}
\begin{center}
\centerline{\psfig{figure=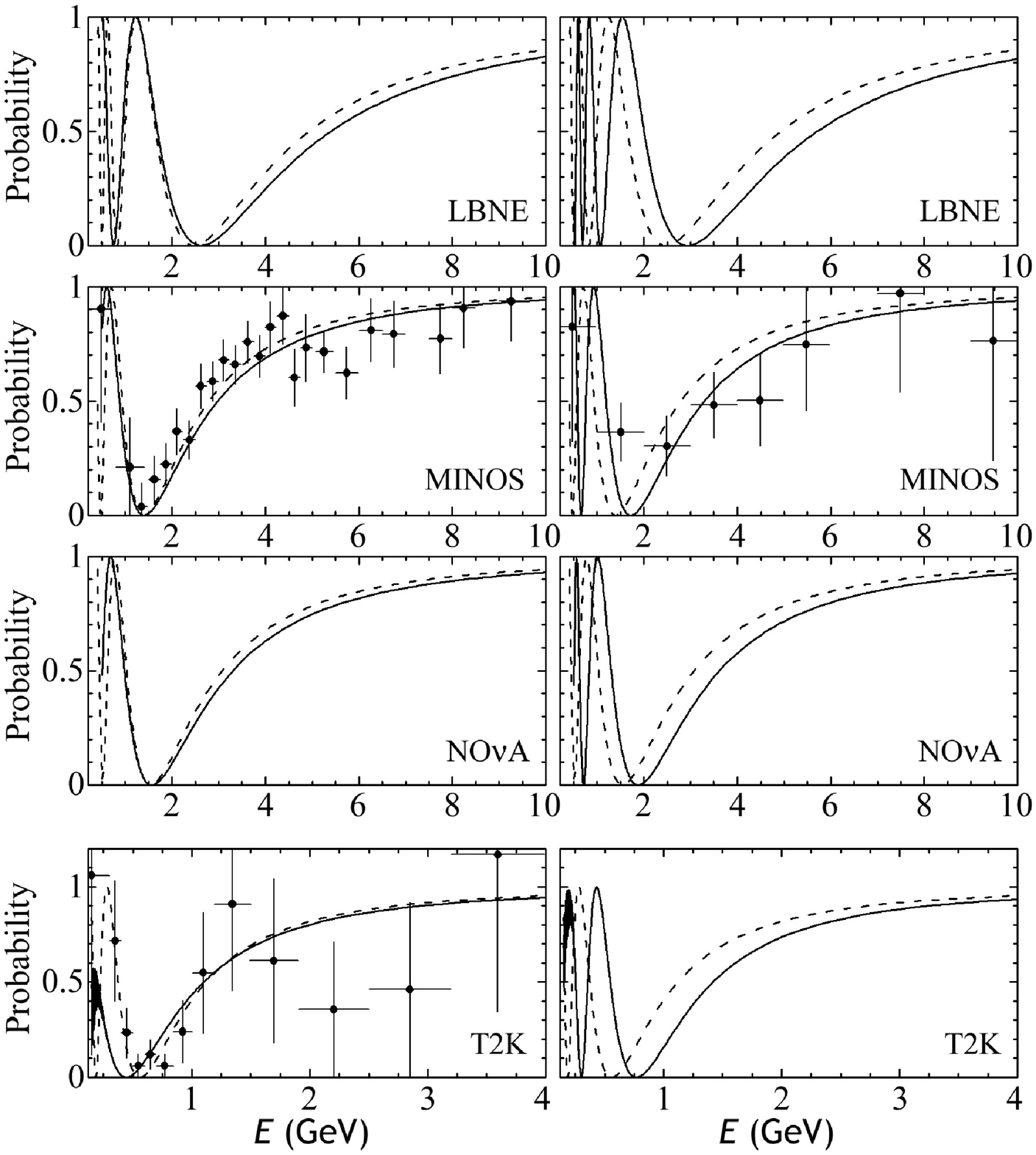,
width=\hsize}}
\caption{ 
Prediction for the probabilities
of $\nu_\mu$ disappearance (left) 
and of $\nub_\mu$ disappearance (right) 
in various long-baseline experiments 
according to the \ceafcfm\ model 
(solid lines; 
$\ch^2_\nu=1.4$, 
$\ch^2_{\ol\nu}=0.9$ for MINOS) 
and the \msm\ 
(dashed lines;
$\ch^2_\nu=1.0$,
$\ch^2_{\ol\nu}=1.6$ for MINOS).
Matter effects are included.
The data are taken 
from Refs.\ \cite{MINOSnu2011,MINOSanomaly,t2kdata}.
}
\label{4coefffig}
\end{center}
\end{figure}

Some indication of a potential difference between
$\nu_\mu$ and $\ol\nu_\mu$ disappearance probabilities
has recently been reported 
by the MINOS collaboration 
\cite{MINOSanomaly}. 
Although the effect may disappear with improved statistics,
a difference of this kind is of interest
in the present context
because it cannot be accommodated in the \msm,
which requires identical neutrino and antineutrino masses. 

In three-parameter puma models, 
the survival probabilities of muon neutrinos and antineutrinos
at MINOS energies are generically the same. 
The result holds because at high energies 
the relevant eigenvalue difference is $\De_{31}\approx 2B^2/C$.
This is even under CPT
provided the $B$ term is either odd or even, 
which is true whenever the $B$ term is only a monomial in $E$.
For the \ceafm\ model
for example,
this symmetry is reflected in Fig.\ \ref{lbnumfig}.
These puma models therefore cannot accommodate
anomalies of the MINOS type either.

In this subsection,
we show that an {\it ad hoc} modification
using an additional parameter
can describe anomalies of this type.
The key idea is as follows.
Instead of choosing a monomial in energy for the $B$ term,
we can take a binomial 
involving two different monomial orders $p$ and $r<p$,
one even and one odd.
This produces a four-coefficient model
with both CPT-odd and CPT-even terms 
contributing at high energies.
If the value of $r$ is close to $p$
and the corresponding coefficients are similar in size,
then the oscillation probabilities 
for neutrinos and antineutrinos differ at high energies. 
In particular, 
the energies of the first oscillation maxima 
of neutrinos and antineutrinos differ,
which is a feature of the MINOS effect.
However,
to preserve compatibility with other experiments,
the $ee$ entry of the effective hamiltonian $\h$ 
must remain unchanged. 
The extra coefficient should therefore appear
only in the $e\mu$ and $e\ta$ entries of $\h$. 
One way to achieve this
is to choose a binomial for the $C$ term as well,
compensating for the modification of the $ee$ entry
arising from the change to $B$.

As an example,
we can add a fourth coefficient to the \ceafm\ model 
while leaving unchanged its main features. 
Choosing $r=4$,
which satisfies the requirement $r<p=5$ 
with $r$ near $p$,
the fourth coefficient can be denoted $\C{4}$.
To generate a neutrino-antineutrino difference at high energies,
it suffices to redefine the $B$ and $C$ terms as 
$B\to B+\C{4}E$, 
$C\to C-\C{4}E$. 
For definiteness,
we can take the numerical values
\bea
m^2 &=& 2.6\times10^{-23} {\rm ~GeV}^2, 
\nn\\
\C{4} &=& 2.0\times10^{-20},
\nn\\
\a{5} &=& -2.6\times10^{-19} {\rm ~GeV}^{-1}, 
\nn\\
\C{8} &=& 1.0\times10^{-16} {\rm ~GeV}^{-4},
\label{c8a5c4m}
\eea
which as before are consistent 
with limits from direct mass measurements,
cosmological mass bounds,
and constraints on anisotropic oscillations. 
These choices preserve the attractive features
of the simpler \ceafm\ model.

The introduction of the fourth coefficient primarily affects 
the probabilities of $\nu_\mu$ and $\nub_\mu$ disappearance 
in long-baseline experiments,
as shown in Fig.\ \ref{4coefffig} 
for the values \rf{c8a5c4m}.
In particular, 
the MINOS data in Ref.\ \cite{MINOSanomaly}
are reasonably described by the model.
The location of the primary minimum for antineutrino oscillations
is at a higher energy than that for neutrinos,
in agreement with the reported effect.
We emphasize that this result is achieved
with a single additional parameter,
without any masses,
in a global model of neutrino oscillations.
Note also that this \ceafcfm\ model predicts a large difference
between the probabilities of $\nu_\mu$ and $\nub_\mu$ disappearance 
in the T2K experiment.
However,
the $\nu_e$ appearance probabilities in long-baseline experiments
are essentially unchanged from those for the \ceafm\ model 
shown in Fig.\ \ref{lbnuefig}.

\subsection{Enhanced models}
\label{Sec: Enhancement}

\begin{figure}
\begin{center}
\centerline{\psfig{figure=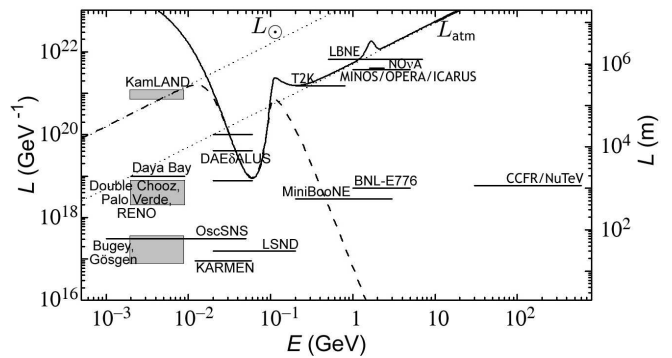,
width=\hsize}}
\caption{ 
Energy dependences of the oscillation lengths
for antineutrinos in the doubly enhanced puma model.
The corresponding plot for neutrinos 
in the top panel of Fig.\ \ref{kmplot}
remains unaffected. 
The disappearance lengths for the model are 
$\ltob$ (solid line) and $\lttb$ (dashed line),
displayed for the values given 
by Eqs.\ \rf{c8a5c4m}, \rf{enhance1}, and \rf{enhance2}.
The dotted lines are the disappearance lengths 
$L_\odot$ (solar) and $L_\text{atm}$ (atmospheric)
in the \msm.
}
\label{enhancedkmfig}
\end{center}
\end{figure}

\begin{figure}
\begin{center}
\centerline{\psfig{figure=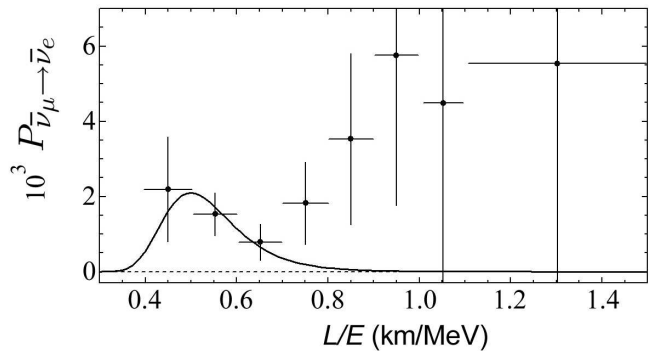,
width=\hsize}}
\caption{ 
Comparison of enhanced puma model 
(solid line, 
$\ch^2=1.6$) 
and the \msm\ 
(dashed line,
$\ch^2=2.6$)
with 
LSND antineutrino data 
\cite{MiniBooNE1}.
}
\label{lsndfig}
\end{center}
\end{figure}

\begin{figure}
\begin{center}
\centerline{\psfig{figure=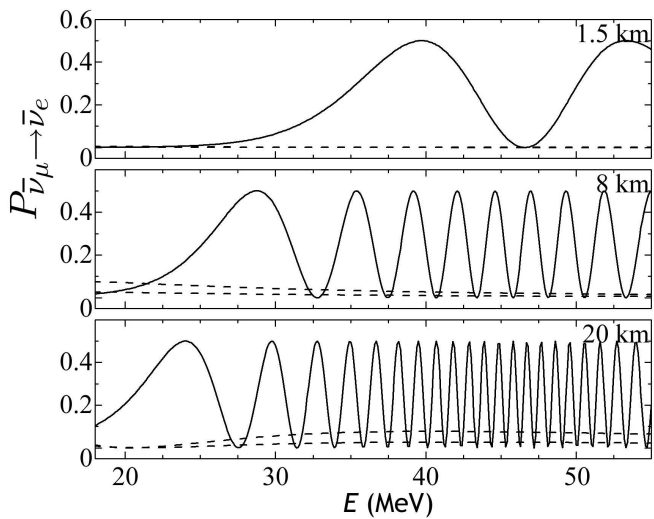,
width=\hsize}}
\caption{ 
Predictions for the probability of $\ol\nu_e$ appearance 
at the three baselines proposed for the DAE$\de$ALUS experiment
according to the enhanced puma model (solid lines) 
and the \msm\ 
(upper dashed lines, $\sin^22\th_{13}=0.15$;
lower dashed lines, $\sin^22\th_{13}=0.05$). 
}
\label{enhanceddaedalusfig}
\end{center}
\end{figure}

\begin{figure}
\begin{center}
\centerline{\psfig{figure=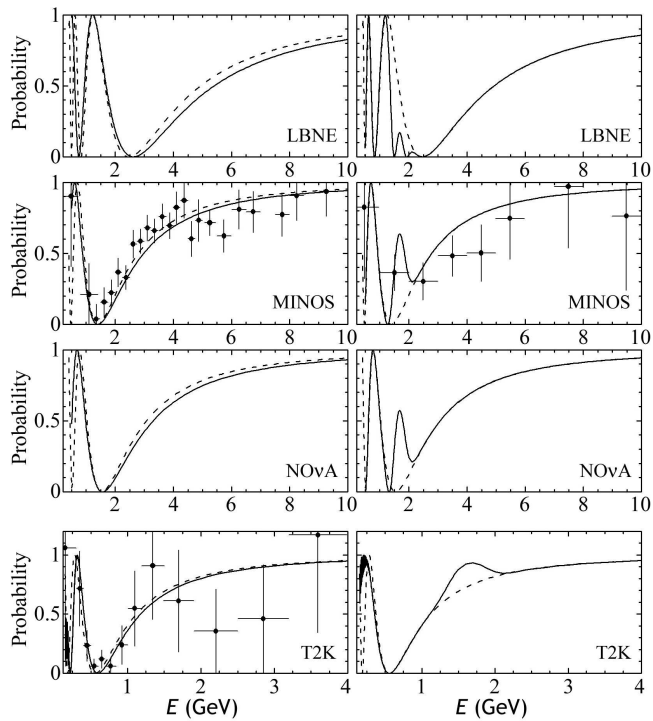,
width=\hsize}}
\caption{ 
Prediction for the probabilities
of $\nu_\mu$ disappearance (left) 
and of $\nub_\mu$ disappearance (right) 
in various long-baseline experiments 
according to the doubly enhanced puma model 
(solid lines; 
$\ch^2_\nu=1.3$, 
$\ch^2_{\ol\nu}=0.9$ for MINOS) 
and the \msm\ 
(dashed lines;
$\ch^2_\nu=1.4$,
$\ch^2_{\ol\nu}=1.8$ for MINOS).
Matter effects are included.
The data are taken 
from Refs.\ \cite{MINOSnu2011,MINOSanomaly,t2kdata}.
}
\label{enhancedpmmfig}
\end{center}
\end{figure}

In a search for appearance of electron antineutrinos
in a beam of muon antineutrinos,
the LSND experiment 
found evidence for a small probability 
$P_{\ol\nu_\mu\to\ol\nu_e}\simeq0.26\pm0.08\%$ 
of $\ol\nu_\mu\to\ol\nu_e$ oscillations 
at baseline $L=30$ m
and energies in the range 20-60 MeV 
\cite{LSND}.
This signal cannot be accommodated within the \msm\
because the required mass-squared difference 
$\De m^2_{\rm LSND}\simeq 1$ eV$^2$
is orders of magnitude larger
than $\De m^2_\odot$ and $\De m^2_\text{atm}$.

In the general puma model,
the oscillation $\nub_\mu\to\nub_e$ is given exactly by
\bea
P_{\nub_\mu\to\nub_e} &=&
8\fr{(A+\ol B)^4}
{\ol N_1^2\ol N_2^2}
\sin^2\left(\fr{\ol\De_{21}L}{2}\right),
\label{Pem(puma)}
\eea
which is governed by the same eigenvalues 
that control reactor antineutrinos.
The energies in the LSND experiment
are greater than those of reactor antineutrinos,
but the appearance length is smaller
by about two orders of magnitude.
Achieving this with a monomial energy dependence in $\h$
while preserving consistency with other experiments is challenging,
as it requires a large power of the energy 
and a seesaw triggered around 10 MeV. 

An interesting option generating 
the required steep fall and rise in $\lttb(E)$ 
is to introduce a smooth nonpolynomial function with a peak.
A function of this type could act as an enhancement of $\hb$
arising from a series of coefficients in the SME.
It can be approximated generically
using at least three parameters,
one to position it,
one to fix its height,
and one to specify its width.
A simple example is a gaussian enhancement of the form
\beq
\de \ol h =\al\exp[-\be(E-\ve)^2].
\label{h_gaussian}
\eeq
To preserve the $S_2$ symmetry of $\hb$,
the enhancement can be limited to 
the $\ol e \hskip 1pt\ol\mu$ 
and $\ol e \hskip 1pt\ol\ta$ 
entries of $\hb$ via the redefinitions
$\ol B\to \ol B+\de \ol h$ and $\ol C\to \ol C-\de \ol h$. 
Under the CPT transformation
mapping $\hb$ to $\h$,
the signs of $\al$ and $\ve$ change.
As a result, 
either the function or its CPT conjugate 
is localized at an unphysical value of the energy,
and so the enhancement affects 
either neutrinos or antineutrinos but not both.

As an example,
consider an enhanced \ceafm\ model
with specific enhancement values
\bea
\al &=& 3.0\times 10^{-19} {\rm ~GeV}, 
\nn\\
\be &=& 3.0\times 10^3 {\rm ~GeV}^{-2}, 
\nn\\
\ve &=& 60 {\rm ~MeV}.
\label{enhance1}
\eea
The positive value of $\ve$ ensures
that only antineutrinos are affected.
This enhancement produces a sharp dip
in the disappearance length $\lttb(E)$
centered around 60 MeV,
as displayed in Fig.\ \ref{enhancedkmfig}. 
The resulting oscillation probability
$P_{\ol\nu_\mu\to\ol\nu_e}$
contains a nonzero signal
in the same region as the LSND data,
as shown in Fig.\ \ref{lsndfig}.
Since the $\lttb$ curve passes through the 
region of sensivity for the DAE$\de$ALUS experiment
\cite{DAEdALUS},
large signals are predicted in all three detectors
as shown in Fig.\ \ref{enhanceddaedalusfig}.
A large oscillation signal is also predicted
at high energies 
in the OscSNS experiment
\cite{oscsns}.

Differences between 
$\nu_\mu$ and $\ol\nu_\mu$ disappearance probabilities
can also be generated by another enhancement 
of this general type.
Two simultaneous enhancements can be included 
without interference
provided they are localized in different regions
of the spectrum. 
Figure \ref{enhancedkmfig}
shows the effect on the disappearance length $\ltob$
of adding a second enhancement with the values 
\bea
\al_2 &=& -2.0\times10^{-19} {\rm ~GeV}, 
\nn\\
\be_2 &=& 13 {\rm ~GeV}^{-2}, 
\nn\\
\ve_2 &=& 1.7 {\rm ~GeV}.
\label{enhance2}
\eea
The resulting disappearance probabilities
for muon neutrinos and antineutrinos
in long-baseline experiments 
are shown in Fig.\ \ref{enhancedpmmfig}.
While the single extra coefficient 
in the four-parameter \ceafcfm\ model is more economical
in generating an anomaly like that reported by MINOS
\cite{MINOSanomaly},
the introduction of the second enhancement 
centered near 2 GeV produces interesting and 
distinctive oscillation signals 
in the LBNE and NO$\nu$A experiments 
for $\nu_\mu$ disappearance, 
as shown in Fig.\ \ref{enhancedpmmfig}.
In contrast,
the $\nu_e$ appearance probabilities 
displayed for the \ceafm\ model in Fig.\ \ref{lbnuefig} 
are largely unaffected by the enhancement.

\section{Discussion}
\label{Discussion}

In this work,
we have investigated the behavior of neutrinos
governed by an effective hamiltonian $\h$
of the puma form \rf{h(puma)}.
This texture is interesting in part because
it leads to descriptions of neutrino oscillations 
that are globally compatible with experimental data.
The associated Lorentz-violating models are intriguing
because they are frugal,
they have a certain elegance,
and their novel features
are compatible with data in unexpected ways.
We remark in passing that the existence 
of these models was unclear {\it a priori},
becoming apparent only through 
a systematic search for viable candidates.

The frugality can be traced 
to the use of only two degrees of freedom 
to describe established data,
instead of the usual five required by the \msm,
while a third degree of freedom
efficiently encompasses the MiniBooNE anomalies.
Adding a fourth degree of freedom
readily generates an anomaly of the MINOS type,
while a three-parameter enhancement
produces a signal in the LSND experiment.
These four latter degrees of freedom are {\it ad hoc},
and their necessity depends on the ultimate confirmation
of the MINOS and LSND anomalies.
However,
to our knowledge
the resulting texture represents 
the sole extant global model 
for neutrino oscillations,
and moreover uses degrees of freedom 
comparable in number to those of the \msm.  

The symmetry of $\h$ also implies a certain elegance.
The puma texture \rf{h(puma)} could naturally stem
from more fundamental physics at the unification scale
that generates a democratic contribution 
to the dominant mass operator 
in the low-energy effective theory.
The resulting $S_3$ symmetry 
then holds at low energies and ensures tribimaximal mixing
but is broken to $S_2$ at higher energies 
by subdominant terms in the SME.
This symmetry structure leads to the 
attractive quadratic calculability of the models.
The coefficients required for compatibility with data
are of plausible Planck-suppressed size.

The novel features of the puma models
originate in the unconventional energy dependence
in the eigenvalues of $\h$
and the mixing matrix $U$.
Indeed,
it is a pleasant surprise that the models pass the test
of compatibility with existing data,
despite their qualitative differences
compared to the \msm.
One striking feature is the Lorentz-violating seesaw,
which makes viable the absence 
of a mass parameter at high energies.
Another satisfying feature is 
the steep drop with energy of the oscillation length $\ltt$,
which is naturally enforced by the third degree of freedom
required to generate the Lorentz-violating seesaw.
As discussed above,
this drop enables $\ltt$ to attain the vicinity
of the MiniBooNE experiment in $E$-$L$ space,
thereby generating a low-energy signal
compatible with the MiniBooNE anomaly. 
Moreover,
this feature appears in conjunction with a rapid decrease 
in the relevant oscillation amplitudes
accompanying the large oscillation phase.
This accounts for null signals 
in high-energy short-baseline accelerator experiments
without invoking the tiny oscillation phase of the \msm.

The puma texture $\h$ predicts certain signals
that differentiate sharply between it and the \msm. 
One key feature is the energy dependence 
of the effective mixing angle $\th_{13}$.
This implies the probability of $\nu_e$ appearance 
is larger in the T2K experiment than in the MINOS experiment.
It also predicts no accompanying signal in reactor experiments,
a result at odds with the \msm. 
With an enhancement present,
strong signals are predicted in experiments
at intermediate energies and moderate baselines
such as the proposed DAE$\de$ALUS experiment.
Another unique signal predicted by some models is CPT violation,
which implies differences in oscillation probabilities
between neutrinos and antineutrinos.
Perhaps the most direct evidence for Lorentz violation
would be the discovery of oscillation anisotropies
arising from the boost relative to the isotropic frame.
One signal would be sidereal variations of oscillations
in the laboratory frame
\cite{ak},
which in the puma models are predicted to be some 10-100 times
below current sensitivities 
\cite{puma}.
In any event,
the results in this work
show that Lorentz- and CPT-violating models 
can serve as an experimentally viable foil to the \msm,
while offering a simple and credible alternative 
for realistic modeling of neutrino oscillations.

\section*{Acknowledgments}

This work was supported in part
by the Department of Energy
under grant DE-FG02-91ER40661
and by the Indiana University Center for Spacetime Symmetries.

\end{document}